\documentclass[11pt]{amsart}
\usepackage{amsaddr}
\usepackage{amssymb}
\usepackage{enumerate}
\usepackage{enumitem}
\usepackage{amsmath,amsthm,marvosym,wasysym,mathdots}
\usepackage{bbm}
\usepackage{natbib}
\usepackage{color}
\usepackage{setspace}
\usepackage{epsfig}
\usepackage{hyperref}
\usepackage{subcaption}
\usepackage{pbox}
\usepackage[stable]{footmisc}
\hypersetup{%
  colorlinks=false,
  pdfborder = {0 0 0.5 [3 0]}
}
\usepackage{breakurl}
\newcommand{\indic}[1]{1\hspace{-2.1mm}{1}_{\{#1\}}} 

\usepackage{caption}
\usepackage{multirow}
\usepackage[left=1in,right=1in,top=1.1in,bottom=1in]{geometry}

\renewcommand{\vec}[1]{\mathbf{#1}} 

\newtheorem{theorem}{Theorem}
\newtheorem*{theorem*}{Theorem}

\newtheorem{definition}[theorem]{Definition}

\newtheorem{lemma}[theorem]{Lemma}

\newtheorem{proposition}[theorem]{Proposition}
\newtheorem{remark}[theorem]{Remark}

\theoremstyle{remark}

\def\N{{\mathbb N}} 
\def\Q{{\mathbb Q}} 
\def\R{{\mathbb R}} 
\def\P{{\mathbb P}} 
\newcommand{\EE}{{\mathord{I\kern -.33em E}}}
\def\E{{\mathbb E}} 

\def\1{1{\hskip -3.3 pt}\hbox{I}}
\def\F{{\mathcal F}}

\def\H{{\mathcal H}}

\providecommand{\varitem}{}


\numberwithin{equation}{section}
\numberwithin{theorem}{section}

\usepackage{float}
\usepackage{hhline}
\usepackage[font=small,skip=0pt]{caption}
\usepackage{bm}
\usepackage{pbox}
\numberwithin{equation}{section}
\numberwithin{theorem}{section}

\usepackage{algorithm} 
\usepackage{algpseudocode} 
\algdef{SE}[DOWHILE]{Do}{doWhile}{\algorithmicdo}[1]{\algorithmicwhile\ #1}%
\algblock{Inputs}{EndInput}
\algnotext{EndInput}
\algblock{Output}{EndOutput}
\algnotext{EndOutput}

\begin{document}

\title[A L\'evy-Driven OU Process for the Valuation of CDX Swaptions]{A L\'evy-Driven Ornstein-Uhlenbeck Process for the Valuation of Credit Index Swaptions}
\author{Yoshihiro Shirai}
\date{\today} 
\email{yshirai@umd.edu}
\address{Department of Mathematics, University of Maryland, College Park}
 
\begin{abstract}
A L\'evy-driven Ornstein-Uhlenbeck process is proposed to model the evolution of the risk-free rate and default intensities for the purpose of evaluating option contracts on a credit index. Time evolution in credit markets is assumed to follow a gamma process in order to reflect the different pace at which credit products are exchanged with respect to that of risk-free debt. Formulas for the characteristic function, zero coupon bonds, moments of the process and its stationary distribution are derived. Numerical experiments showing convergence of standard numerical methods for the valuation PIDE to analytical and Montecarlo solutions are shown. Calibration to market prices of options on a credit index is performed, and model and market implied summary statistics for the underlying credit spreads are estimated and compared.
\end{abstract}

\maketitle

\section{Introduction}
This paper proposes a new valuation method for credit index swaptions (``CDXOs'', or ``CDX swaptions''), which are options to enter at a predetermined date a credit index swap. The current literature focuses on single name derivatives (\cite{Jamishidian}, \cite{KrekelWenzel}), or on developing a Black-type formula to retrieve a CDXO price from its quote (typically expressed in terms of the underlying spread), and, particularly, to include the ``front end protection'' into the CDXO payoff (see \cite{BrigoMorini} and \cite{ArmstrongRutkowski}, as well as \cite{Pedersen} and \cite{DoctorGoulden}).\footnote{Applying the conversion formula requires several inputs, such as the CDXO annuity, which are typically unavailable to outsiders. Testing and calibration of the model here proposed with real market data is made possible thanks to time series of CDXO prices provided by Morgan Stanley.} Apart from this formulation and to the best of the author's knowledge, there are no standard valuation methods for credit index derivatives that are consistent with observed credit spreads statistics.

The main contribution of this paper is then to specify an underlying Markov process $\{\vec{X}_t\}_{t\geq 0}$ that ultimately defines both short rate and credit spread dynamics and is such that:
\begin{itemize}
[noitemsep,nolistsep]
\item[(i)] a reliable and fast numerical method can be implemented to obtain CDX and CDXO prices;
\item[(ii)] the model parameters can be calibrated to fit well the option market price surface; and
\item[(iii)] the model implied statistical properties of the credit spread fit those implied by the market.
\end{itemize}
We assume in particular that $\{\vec{X}_t\}_{t\geq 0}$ is the two dimensional process whose components $\{r_t\}_{t\geq 0}$ and $\{\lambda_t\}_{t\geq 0}$ represent, respectively, the short rate, and the default intensity process of each entity in the underlying index. The default time for entity $i$ is then modeled as the first time the default intensity integrated process $\{\Lambda_t\}_{t\geq 0}$ reaches a threshold $\varepsilon_i$, where $\varepsilon_1,...,\varepsilon_n$ are independent copies of an exponential random variable and $n$ is the numbers of entities in the index.

We take mean reverting processes for $\{r_t\}_{t\geq 0}$ and $\{\lambda_t\}_{t\geq 0}$. Randomness of $\{r_t\}_{t\geq 0}$ is represented by a gamma process $\{g^r_t\}_{t\geq 0}$, whereas for $\{\lambda_t\}_{t\geq 0}$ it is the sum of a double gamma process $\{\tilde{g}^\lambda_t\}_{t\geq 0}$ and a scalar multiple $\rho$ of $\{g^r_t\}_{t\geq 0}$ itself. Multiple gamma processes were first investigated in \cite{MadanEntropy}, for the purpose of randomizing the average speed of trading activity.

The distribution of the increments of the process $\{\vec{X}_t\}_{t\geq 0}$ is self-decomposable (see \cite{Sato1984}), i.e. the increments of $\{\vec{X}_t\}_{t\geq 0}$ are scaled sums of independent random variables. As noted in \cite{CGMY}), this reflects the fact that financial variables, including default intensities, are determined by a large number of random news. The experiments reported in this paper is one of the few study of the goodness of fit of a self-decomposable law to CDX data. The paper is also motivated by the need of a purely discontinuous process to utilize the theory of Conic Finance (\cite{MadanPistoriusStadje}) to evaluate credit products. Because of this, and although the exploration of the applications of Conic Finance to credit derivatives is left to future research, we ignore here the relatively small accounting issues related to the front end protection. 

Default times defined as above, known as doubly stochastic random times, are commonly used in credit risk modeling (see \cite{Bielecki2002}, \cite{JeanblancRutkowski} and \cite{QRM}) and their development goes back to \cite{Duffie1999}, \cite{Lando1998}, \cite{JarrowLandoTurnbull97} and \cite{Madan1998}. Common specifications for rate and intensity processes are the affine models developed in \cite{DuffieKan} for diffusion models and in \cite{Duffie2000} and \cite{duffiegarleanu} for affine jump-diffusion models. Reduced form model with non doubly stochastic random times are also possible, although such a direction is not investigated here. For their development the reader is referred to \cite{Kusuoka1999} and \cite{ElliottJeanblancYor}.

We derive the infinitesimal generator of the process $\{\vec{X}_t\}_{t\geq 0}$, based on which prices of zero coupon bonds can be computed analytically. Moments and characteristic exponent of $\vec{X}_t$ for $t\in [0,\infty]$ are also computed analytically, and level curves of its bivariate density for different parameters are plotted using a 2D-version of the FFT algorithm (similar to the one in \cite{Hurd}). We then derive analytical formulas for discounted payoffs of credit index swaps, and the partial integro differential equation (PIDE) for credit index swaptions prices, together with a finite difference scheme for its solution. Calibration is performed for each maturity to all traded strikes of options on the IG CDX index as of 2 January 2020. We do not perform a stability analysis, but we show that the numerical error for a given set of parameters (obtained from calibration to market prices on 2 January 2020) is close enough to the prices obtained via Montecarlo simulation. 

Finally, we compare market and model implied summary statistics of credit spreads for a specific maturity. As shown in \cite{CarrMadan2001}, variance, skewness and kurtosis of an equity position under the risk neutral measure can be replicated with a continuum of option contracts. Here it is shown that variance, skewness and kurtosis of the spread of a credit index can be replicated with a continuum of credit index swaptions under the measure $\Q^{A}$ corresponding to choosing as numeraire the annuity of the index. Our model is then calibrated to market prices and market and model implied variance, skewness and kurtosis of the credit spreads are compared. The results of our analysis show that our model can capture positive skewness and leptokurtic features of CDX spreads under the measure $\Q^A$, and the model and market implied moments are of the same magnitude. We observe, in particular, that the replication of credits spreads with option contracts under the measure $\Q^A$ represents a new way to extract model-free statistical properties of credit spreads from market prices of options.

The rest of the paper is organized as follows. In section 2 we review the basics of credit index derivatives and their market, and in section 3 the fundamental mathematical framework is introduced. In section 4 we specify the pure-jump dynamics of short rate and default intensity, derive the characteristic exponent of the underlying Markov process and the valuation PIDE. A simple finite difference scheme is tested in section 5, and a comparison of model and market results is shown in section 6. Section 7 concludes.

\section{Credit Index Derivatives and their Market}
The last few decades saw a spectacular rise in trading volumes of credit derivatives, such as credit default swaps (``CDS''), credit index swaps (``CDX''), single tranche credit default obligations, etc. One reason for this is that the main features of CDS contracts, which form the basic asset class in credit markets, have been standardized,\footnote{For instance, banks and financial institutions typically utilize the ISDA Master Service Agreement, published by the International Swaps and Derivatives Association, as the framework agreement such that each futures transactions between the parties of the agreement are mostly defined by it, leaving only specific points of the transaction open to negotiation.} thus allowing a relatively easy implementation of hedging and speculative strategies and increasing the liquidity of CDS market with respect to that of corporate bonds. However, the details of credit derivatives contracts remain complex and satisfactory valuation methods for the case of CDX forwards and swaptions are yet to be determined.

To introduce the mathematical problem, recall that a CDS is an over the counter contract between two counterparties - the protection buyer and seller - in which protection against the risk of default of an underlying entity (usually a company issuing bonds in the debt market) is provided by the seller to the buyer. The latter pays the former a predetermined premium $K$ (defined as a credit spread multiplied by the contract's notional) at regular intervals until the contract expires and obtains a contingent payment from the seller triggered by any credit event (such as default, restructuring, downgrade, etc.) concerning the underlying entity.

A CDX can be thought of as a portfolio of credit default swaps. There are two families of credit indices: CDX.NA and iTraxx. CDX.NA indices refer to North American companies and iTraxx indices refer either to European firms or to Asian and Australian ones. Each family is composed of different indices, each of which representing a different class of credit quality. A summary of the main credit indices is shown in table \ref{table:CDX}. It is important to observe that, in order to reflect changes in the credit quality of the constituents, the composition of most credit indices changes every six months on March 20 and September 20. Each series of an index corresponds to a specific roll date, and older series continue to trade, but their market is far less liquid (see \cite{QRM}).

\begin{table}[H]
\begin{center}
  \begin{tabular}{ c | c | c | c }
	Name & Pool size & Region & Credit Quality\\
	\hline
    CDX.NA.IG & 125 & North America & Investment Grade\\
    CDX.NA.IG.HVOL & 30 & North America 
    	& Low-quality Investment Grade\\
    CDX.NA.HY & 100 & North America & Speculative Grade\\
    iTraxx Europe & 125 & Europe & Investment Grade\\
    iTraxx Europe & 30 & Europe & Low-quality Investment Grade\\
    \hline
  \end{tabular}
\end{center}
\caption{Major credit indices and their characteristics (source: \cite{QRM}).}
\label{table:CDX}
\end{table}

Similarly to a CDS, the cash flow associated to a CDX consists again of a premium payment leg (with payments made by the protection buyer) and a default payment leg (with payments made by the protection seller). Premium payments, which are defined as a credit spread multiplied by the index annuity (a measure of the number of underlying issuers for which a credit event has not occurred yet) are due at deterministic dates $T_0<T_1<...<T_M$, where $T_M$ is the maturity of the contract and $T_0$ the inception date (for forward-start contracts $T_0>0$). A credit event concerning any of the underlying entities triggers a payment by the seller. Standardized CDXs have quarterly premium payments and maturity at issuance is three, five, seven and ten years, with five years being the most liquid traded maturity. 

There are two main differences between a CDX and a (portfolio of) CDS: (1) the contingent payment of a CDX is the same for each underlying entity and (2) it does not become an empty contract after a single credit event occurs, so the expected discounted value of the cumulated losses before the inception date (i.e. the above mentioned front end protection) is included in the price. 

\section{Review and Assumptions}

\subsection{Hazard Rates and Doubly Stochastic Random Times}
\begin{definition} Suppose that:
\begin{itemize}
\item[i.] $(\Omega,\F,\Q)$ is a probability space;
\item[ii.] $\{\F_t\}_{t\geq 0}$ is a filtration on $(\Omega,\F,\Q)$ with $\F_{\infty} \subsetneqq \F$ and $\F_t\neq \{\emptyset,\Omega\}$ for all $t\geq 0$;\footnote{Typically $\{\F_t\}_{t\geq 0}$ is generated by a process $\Psi$ measuring economic activity. }
\item[iii.] $\tau:\Omega\rightarrow [0,\infty]$ is $\F$-measurable and $\{\H_t\}_{t\geq 0}:=\{\sigma\left(\{\indic{\tau\leq u}\}_{u\leq t}\right)\}_{t\geq 0}$;
\item[iv.] $\{\Lambda_t\}_{t\geq 0}:=\{\log\left(\Q(\tau> t|\F_{\infty})\right)\}_{t\geq 0}$ is strictly increasing, finite (i.e. $\Q(\tau> t|\F_{\infty})>0$ a.s. for every $t>0$), $\{\F_t\}_{t\geq 0}$-adapted and absolutely continuous, with $\Lambda_t=\int_0^t\lambda_sds$ for $t\geq 0$.
\end{itemize}
Then, $\tau$ is called a doubly stochastic random time with $\F_t$-conditional hazard rate process $\lambda$.
\end{definition}
\begin{remark}
Since $\{\Lambda_t\}_{t\geq 0}$ is $\{\F_t\}_{t\geq 0}$-adapted, we have $\Q(\tau\leq t|\F_t)=\Q(\tau\leq t|\F_\infty)$ $\forall t\geq 0$.
\end{remark}
\begin{lemma}\label{exponentialdoubly} Let $X$ be a standard exponentially distributed random variable on $(\Omega,\F,\Q)$ independent of $\F_{\infty}$, i.e. $\Q(X\leq t|\F_{\infty})=1-e^{-t}$ for every $t\geq 0$. Let $\{\lambda_t\}_{t\geq 0}$ be a positive $\{\F_t\}_{t\geq 0}$-adapted process such that the process defined by $\Lambda_t=\int_0^t\lambda_sds$ for $t\geq 0$ is increasing and finite. Then, $\tau:=\inf\{t\geq 0:\Lambda_t\geq X\}$ is a doubly stochastic random time with hazard rate process $\{\lambda_t\}_{t\geq 0}$.
\end{lemma}
\begin{proof}
By definition $\{\tau>t\}=\{\Lambda_t<X\}$. Since $\{\Lambda_t\}_{t\geq 0}$ is $\F_{\infty}$-measurable and $X$ is independent of $\F_{\infty}$, we have $\Q(\tau>t|\F_{\infty})
=\Q(\Lambda_t<X|\F_{\infty})=e^{-\Lambda_t}$ for every $t\geq 0$, which proves the result.
\end{proof}
\begin{proposition}[\textbf{Dellacherie's Formulas\footnote{The reference to Dellacherie appears in \cite{JeanblancRutkowski}.}}]\label{ExchangingFiltration}
Let $(\Omega,\F,\Q)$ be a probability space and $\tau$ a doubly stochastic random time with $\{\F_t\}_{t\geq 0}$-conditional hazard rate process $\{\lambda_t\}_{t\geq 0}$. Let $\{r_t\}_{t\geq 0}$ be an $\{\F_t\}_{t\geq 0}$-adapted random process. Suppose that, for some $T>0$, $Y$ is $\F_T$-measurable, $\{\nu_t\}_{0\leq t\leq T}$ and $\{Z_t\}_{t\geq 0}$ are $\{\F_t\}_{t\geq 0}$-adapted.\footnote{Typically, $Y$ is a \textit{survival claim}, i.e. a promised payment if there is no default, $\nu$ is a risky stream of payments that stops when default occurs, and $Z$ is a payment made at default.} Suppose that the random variables
\begin{align*}
|Y|e^{-\int_t^Tr_sds}, \ \int_t^T\nu_se^{-\int_t^sr_udu}ds, \ 
\int_t^T|Z_s\lambda_s|e^{-\int_t^s(r_u+\lambda_u)du}ds
\end{align*}
are integrable with respect to $\Q$ for every $t\geq 0$. Then, for every $t\geq 0$,
\begin{align*}
& \E\left[\left.e^{-\int_t^Tr_sds}
	\indic{\tau>T}Y\right\rvert\F_t\vee \H_t\right]
=	\indic{\tau>t}
	\E\left[\left.e^{-\int_t^T(r_s+\lambda_s)ds}Y
	\right\rvert\F_t\right],\\
& \E\left[\left.\int_t^T\nu_se^{-\int_t^sr_udu}
	\indic{\tau>s}ds\right\rvert\F_t\vee\H_t\right]
=	\indic{\tau>t}
	\E\left[\left.\int_t^T\nu_s
	e^{-\int_t^s(r_u+\lambda_u)du}ds
	\right\rvert\F_t\right],\\
& \E\left[\left.e^{-\int_t^{\tau}r_sds}
	\indic{t<\tau\leq T}Z_{\tau}\right\rvert\F_t\vee\H_t\right]
=	\indic{\tau>t}
	\E\left[\left.\int_t^TZ_s\lambda_s
	e^{-\int_t^s(r_u+\lambda_u)du}ds
	\right\rvert\F_t\right],
\end{align*}
where $\{\H_t\}_{t\geq 0}:=\{\sigma\left(\{\indic{\tau\leq u}\}_{u\leq t}\right)\}_{t\geq 0}$.
\end{proposition}
\begin{proof}
See \cite{QRM}, Theorem 9.23.
\end{proof}

\subsection{Basics of Forward CDS and CDX Contracts}

Consider a forward-start CDS with inception date $T_0$, tenor structure $T_0<...<T_M$, CDS spread $c$ and for a notional of $1$ U.S. dollar. Let $(\Omega,\F,\Q)$ be a probability space, $\{\F_t\}_{t\geq 0}$ a filtration on it, $\{r_t\}_{t\geq 0}$ an $\{\F_t\}_{t\geq 0}$-adapted random process, and $\tau:\Omega\rightarrow [0,\infty]$ a doubly stochastic random time with hazard rate $\{\lambda_t\}_{t\geq 0}$. Assuming that $\tau$ represents the time of the credit event, the payment made by the protection seller (protection leg) discounted at time $t\leq T_0$ is given by
\begin{align*}
\Phi_t
=\delta_{\tau}
 e^{-\int_t^{\tau}r_sds}\indic{T_0<\tau\leq T_M},
\end{align*}
where $\{\delta_t\}_{t\geq 0}$ is the $\{\F_t\}_{t\geq 0}$-adapted process representing loss given default. Similarly, the premium leg is given by
\begin{align*}
\Psi_t=c\sum_{j=1}^Me^{-\int_t^{T_j}r_sds}\indic{\tau>T_j}
[T_j-T_{j-1}]
\end{align*}
Using Dellacherie's formulas, we have
\begin{align*}
\E^{\Q}[\Phi_t|\F_t\vee \mathcal{H}_t]
& = \E^{\Q}\left[\delta_{\tau}e^{-\int_t^{\tau}r_sds}\left(\indic{t<\tau\leq T_M}
	-\indic{t<\tau<T_0}\right)|\F_t\vee \mathcal{H}_t\right]\\
& = \indic{\tau>t}\E^{\Q}\left[\int_{T_0}^{T_M}\lambda_s	
	\delta_se^{-\int_t^s(r_u+\lambda_u)du}ds)|\F_t\right]
\end{align*}
The present value of the protection buyer's cash flow is then given by
\begin{align*}
\E^{\Q}\left[\Phi_t-\Psi_t|\F_t\vee \mathcal{H}_t\right]
& = \indic{\tau>t}\E^{\Q}\left[\int_{T_0}^{T_M}\lambda_s
	\delta_se^{-\int_t^s(r_u+\lambda_u)du}ds)|\F_t\right]\\
& \ \ \
	-\indic{\tau>t}c\sum_{j=1}^M(T_j-T_{j-1})
	\E^{\Q}\left[e^{-\int_t^{T_j}(r_u+\lambda_u)du}|\F_t\right].
\end{align*}
Since the spread $c(t,T_0,T_M)$ on a CDS initialized at time $t$ is chosen such that the time $t$ value of the contract is zero, we then have, on $\{\tau>t\}$,
\begin{align*}
c(t,T_0,T_M)=
\frac{\E^{\Q}\left[
	\int_{T_0}^{T_M}\lambda_s\delta_s
	e^{-\int_t^s(r_u+\lambda_u)du}ds)|\F_t\right]}
{\sum_{j=1}^M(T_j-T_{j-1})
	\E^{\Q}\left[e^{-\int_t^{T_j}(r_u+\lambda_u)du}|\F_t\right]}.
\end{align*}

We next provide the relevant definitions for forward contracts on a CDX (see \cite{BrigoMorini} for details). Suppose that the premium payments occur at $T_0<T_1<...<T_M$, where $T_M$ is the maturity of the contract and $T_0$ is the inception date. Define the following quantities:
\begin{itemize}
\item[(i)] \textbf{Cumulated losses}: $L_t=\frac{\delta}{n}\sum_{j=1}^n\indic{\tau_j<t}$, where $\delta$ is the loss given default (typically common for each name and nonrandom) and $\tau_j$ is the time of default of entity $j$;
\item[(ii)] \textbf{Outstanding notional}: $N_t=1-\frac{L_t}{\delta}$;
\item[(iii)] \textbf{Index Annuity}:
\begin{align*}
A_t=\sum_{j=1}^Me^{-\int_t^{T_j}r_udu}
	\int_{T_{j-1}}^{T_j}N_sds
	\approx \sum_{j=1}^{M}e^{-\int_t^{T_j}r_udu}
	N(T_j)(T_j-T_{j-1});
\end{align*}
\item[(iv)] \textbf{Premium leg}: $\Psi(t,c)=cA_t$;
\item[(v)] \textbf{Protection leg}: 
\begin{align*}
\Phi_t=\int_{T_0}^{T_M}e^{-\int_t^{s}r_udu}dL_s
	\approx \sum_{j=1}^{M}e^{-\int_t^{T_j}r_udu}
		\left[L(T_j)-L(T_{j-1})\right]
\end{align*}
\item[(vi)] \textbf{Front End Protection}: $F_t=e^{-\int_t^{T_0}r_sds}L_{T_0}$, $t\leq T_0$, is the discounted value at time $t$ of cumulated losses at time $T_0$.
\end{itemize}
The discounted payoff of a CDX is then given by
\begin{align}\label{DiscountedPayoffCDXExact}
e^{-\int_t^{T_0}r_sds}\left[\Phi_{T_0}-\Psi_{T_0}+F_{T_0}\right]
=\Phi_t-\Psi_t+F_t.
\end{align}
Thus, any time a default event is triggered for any of the names composing the index, the name that defaulted is removed from the index and a payment of size $\delta/n$ is made by the protection seller, provided the default event happens after the inception date of the CDX. If the event happens before the inception date, then the name that defaulted is again removed from the index, and the loss is paid at the inception of the swap.

As mentioned, we avoid technical complications related to the front end protection, and assume no defaults occur before inception. The discounted payoff at $0\leq t\leq T_0$ is then
\begin{align}\label{DiscountedPayoffCDX}
e^{-\int_t^{T_0}r_sds}\left[\Phi_{T_0}-\Psi_{T_0}\right]
=\Phi_t-\Psi_t.
\end{align}
To model different default rates, we consider a common intensity process $\{\lambda_t\}_{t\geq 0}$ for each underlying name (i.e. the pool is ``homogeneous''), and, for $i=1,...,n$, we define the default time
\begin{align*}
\tau^i=\inf\left\lbrace t>0:\Lambda_t>\varepsilon_i\right\rbrace,
\end{align*}
where $\varepsilon_1,...,\varepsilon_n$ are independent exponential random variables that are also independent of $\mathcal{F}_{\infty}$. Then (\cite{QRM}, lemma 9.33), the default times are conditionally independent doubly stochastic random times, i.e. each $\tau^i$ is a doubly stochastic random time with respect to $\F_t$ and
\begin{align*}
\Q(\tau^1>t,...,\tau^n>t|\F_{\infty})
=\prod_{i=1}^n\Q(\tau^i>t|\F_{\infty}).
\end{align*}
In this case, we obtain for $i=1,...,n$,
\begin{align*}
\E^{\Q}\left[e^{-\int_{T_0}^{T_\ell}r_udu}
	\indic{\tau^{i}>T_{\ell-1}}|
	\F_{T_0}\vee \mathcal{H}_{T_0}\right]
& = \E^{\Q}\left[e^{-\int_{T_0}^{T_\ell}r_udu}
	\indic{\tau^{i}>T_{\ell-1}}|
	\F_{T_0}\vee \mathcal{H}^i_{T_0}\right],
\end{align*}
where, $\forall T\geq t\geq 0$, $\mathcal{H}_{t}:=\vee_{i=1}^n\mathcal{H}^i_{t}$ and $\mathcal{H}^i_t:=\sigma(\{\indic{\tau^i\leq u}\}_{u\leq t})$. From, (1), the Dellacherie's formulas given in Proposition \ref{ExchangingFiltration}, (2), the tower property of conditional expectation and assuming, (3), that the vector process $\{\vec{X}_t\}_{t\geq 0}$ composed of the short rate and default intensity processes is Markovian, it follows that there is $g_{\ell}:\R_+^3\rightarrow \R$ for $1\leq \ell\leq M$, such that, for $1\leq i\leq n$,
\begin{align*}
\E^{\Q}\left[e^{-\int_{T_0}^{T_\ell}r_udu}
	\indic{\tau^{i}>T_{\ell-1}}|
	\F_{T_0}\vee \mathcal{H}_{T_0}\right]
& = \E^{\Q}\left[e^{-\int_{T_0}^{T_{\ell-1}}r_udu}
	\indic{\tau^{i}>T_{\ell-1}}
	P(T_{\ell-1},T_\ell)|
	\F_{T_0}\vee \mathcal{H}_{T_0}\right]\\
& = \indic{\tau^{i}>T_0}
 	\E^{\Q}\left[e^{-\int_{T_0}^{T_{\ell-1}}(r_u+\lambda_u)du}
	P(T_{\ell-1},T_\ell)|\F_{T_0}\right]\\
& = \indic{\tau^{i}>T_0}g_{\ell}(T_0,r_{T_0},\lambda_{T_0}),
\end{align*}
where, for every $T\geq t\geq 0$,
\begin{align*}
P(t,T)=\E^{\Q}\left[e^{-\int_t^Tr_udu}\rvert \F_t\right].
\end{align*}
Similarly, there is a function $h_{\ell}:\R_+^3\rightarrow \R$ such that
\begin{align*}
\E^{\Q}\left[e^{-\int_{T_0}^{T_\ell}r_udu}
	\indic{\tau^{i}>T_{\ell}}|
	\F_{T_0}\vee \mathcal{H}_{T_0}\right]
& = \indic{\tau^{i}>T_0} 
 	\E^{\Q}\left[e^{-\int_{T_0}^{T_{\ell}}(r_u+\lambda_u)du}
	|\F_{T_0}\right] 
  = \indic{\tau^{i}>T_0}h_{\ell}(T_0,r_{T_0},\lambda_{T_0}).
\end{align*}
Therefore, using the tower property of conditional expectation, we have
\begin{align*}
\E^{\Q}\left[
	\Phi_{T_0}|\F_{T_0}\vee
	\mathcal{H}_{T_0}\right]
& = \sum_{\ell=1}^M\frac{\delta}{n}\sum_{i=1}^n
	\E^{\Q}\left[e^{-\int_{T_0}^{T_\ell}r_udu}
	(\indic{\tau^{i}>T_{\ell-1}}-\indic{\tau^{i}>T_{\ell}})	
	|\F_{T_0}\vee\mathcal{H}_{T_0}\right]\\
& = \frac{\delta}{n}\sum_{i=1}^n
    \indic{\tau^{i}>T_0}\left[
	(g(T_0,r_{T_0},\lambda_{T_0})
	-h(T_0,r_{T_0},\lambda_{T_0}))\right],
\end{align*}
where $g:=\sum_{\ell=1}^Mg_\ell$, $h:=\sum_{\ell=1}^Mh_\ell$.

As for the premium leg, similar calculations give
\begin{align*}
\E^{\Q}\left[A_{T_0}|\F_{T_0}\vee
	\mathcal{H}_{T_0}\right]
& = \sum_{\ell=1}^M(T_\ell-T_{\ell-1})\frac{1}{n}\sum_{i=1}^n
	\E^{\Q}\left[e^{-\int_{T_0}^{T_\ell}r_udu}
	\indic{\tau^{i}>T_\ell}|\F_{T_0}\vee
	\mathcal{H}_{T_0}\right]\\
& = \sum_{\ell=1}^M(T_\ell-T_{\ell-1})
	\frac{1}{n}\sum_{i=1}^n
	\indic{\tau^{i}>T_0}h_{\ell}(T_0,r_{T_0},\lambda_{T_0})
\end{align*}
The payoff $\pi(T_0,r_{T_0},\lambda_{T_0})$ of a receiver CDX swaption\footnote{An option contract on a CDX index is of receiver type if the holder has the right, not the obligation, to sell protection, and it is of payer type if the holder has the right, not the obligation, to buy protection.} is then
\begin{align*}
&\pi(T_0,r_{T_0},\lambda_{T_0})\\
& \ \ \
 =\left[
 	\kappa\sum_{\ell=1}^M(T_\ell-T_{\ell-1})
	h_{\ell}(T_0,r_{T_0},\lambda_{T_0})
	-\delta\left(g(T_0,r_{T_0},\lambda_{T_0})
	-h(T_0,r_{T_0},\lambda_{T_0})\right)
	\right]^+.
\end{align*}
This allows one to define a PIDE for the price of a CDX swaption at time $t$ in terms of the vector $(t,r_t,\lambda_t)$.\footnote{The value of $\lambda_t$ is not directly observable, but can be retrieved by equating to zero the value of the forward contract at the strike for which put-call parity holds.} Observe that, since the price of a CDX swaption initiated at time $t$ is zero at time $t$, the corresponding credit spread, denoted by $c(t,r_t,\lambda_t,T_0,T_M)$, is given by
\begin{align}\label{CDXSpreadFormula}
c(t,\lambda_t,T_0,T_M)
&:=\delta\frac{\E^{\Q}\left[\Phi_t|\F_t\right]}
	{\E^{\Q}\left[A_t|\F_t\right]}
  = \delta\frac{g(t,r_t,\lambda_t)-h(t,r_t,\lambda_t)}
	{\sum_{\ell=1}^M(T_\ell-T_{\ell-1})h_\ell(t,r_t,\lambda_t)}.
\end{align}

\section{The Joint Dynamics of the Short Rate and the Default Intensity}
In this section, the dynamics of the process $\{\vec{X}_t\}_{t\geq 0}$ is specified for the purpose of pricing CDX derivatives. Recall that a process $\{g_t\}_{t\geq 0}$ is a gamma process if it is a pure jump Levy process with L\'evy density $\varphi^g$ given by
\begin{align*}
\varphi^g(x)=\gamma\frac{e^{-cx}}{x},
\end{align*}
where $c$ and $\gamma$ are referred to as the scale and shape parameters of the process $\{g_t\}_{t\geq 0}$. Since
\begin{align*}
\log\left(\E^{\Q}\left[e^{iug_t}\right]\right)
=-\gamma t\log\left(1-\frac{iu}{c}\right),
\end{align*}
$g_1$ is a gamma random variable with mean $\tfrac{\gamma}{c}$ and variance $\tfrac{\gamma}{c^2}$. Inspired by \cite{EMPY}, a simple choice for the dynamics of $\{\vec{X}_t\}_{t\geq 0}$ would be to set
\begin{equation}\label{BasicGDSRDI}
d\vec{X}_t
= -\vec{\Theta}^{\ast}\vec{X}_tdt+d\vec{g}_t
\end{equation}
where the asterisk denotes transposition and, for every $t\geq 0$,
\begin{align*}
\vec{\Theta} := \begin{bmatrix}
\theta_r & 0\\
0 & \theta_{\lambda}
\end{bmatrix}, \
\vec{g}_t := \begin{bmatrix}
g^r_t\\
\rho g^r_t+g^{\lambda}_t
\end{bmatrix}
\end{align*}
and where $\{g^r_t\}_{t\geq 0}$ and $\{g^{\lambda}_t\}_{t\geq 0}$ are two independent gamma processes with scale parameters $c_r$ and $c_\lambda$, and shape parameters $\gamma_r$ and $\gamma_{\lambda}$ respectively. Parameters $\theta_r$ and $\theta_{\lambda}$ are positive and measure the speed of reversion toward the (zero) long term average. The impact of jumps in the short rate on the default intensity is modeled by the variable $\rho\geq 0$, assumed nonnegative to ensure a nonnegative default intensity. The initial state $\vec{X}_0$ of the process $\{\vec{X}_t\}_{t\geq 0}$ is assumed independent of $\{\vec{g}_t\}_{t\geq 0}$. Finally, since $\{\vec{g}_t\}_{t\geq 0}$ is a linear transformation of a Levy process, it is itself a Levy process, and $\{\vec{X}_t\}_{t\geq 0}$ is an Ornstein-Uhlenbeck (``OU'') process driven by the L\'evy process $\{\vec{g}_t\}_{t\geq 0}$.

To better fit the option price surface,\footnote{Although not reported here, we experimented with real market data and found that model (\ref{BasicGDSRDI}) is not rich enough to fit, in particular, the prices of out of the money options. Other modeling choices for the default intensity, not explored here, are possible, e.g. one could consider an integrated truncated bilateral gamma process.} and recognizing that economic activity in highly liquid risk-free debt markets (such as the U.S. Treasury market) typically evolves at a different pace than defaultable bonds markets, we subordinate $\{g^{\lambda}_t\}_{t\geq 0}$ to a gamma time change. Specifically, given a third gamma process $\{g^{\tau}_t\}_{t\geq 0}$ with parameters $c_{\tau}$ and $\gamma_{\tau}$, we set, for $t\geq 0$, $\tilde{g}^{\lambda}_t:=g^{\lambda}_{g^{\tau}_t}$, and consider
\begin{equation}\label{GDSRDI}
d\vec{X}_t
= -\vec{\Theta}^{\ast}\vec{X}_tdt+d\tilde{\vec{g}}_t,
\end{equation}
where, for every $t\geq 0$,
\begin{align*}
\tilde{\vec{g}}_t := \begin{bmatrix}
g^r_t\\
\rho g^r_t+\tilde{g}^{\lambda}_t
\end{bmatrix}.
\end{align*}
By \cite{Sato}, Theorem 30.1, $\{\tilde{g}^{\lambda}_t\}_{t\geq 0}$ is a Levy process whose Levy measure satisfies 
\begin{align}\label{LevyMeasure}
\int_0^{\infty}\int_Bp_x(y)dy\varphi_{\tau}(x)dx,
\end{align}
for a Borel set $B\subset \R\setminus\{0\}$. Here, $p_x$ is the density of a Gamma distribution with parameters $c_{\lambda}$ and $\gamma_{\lambda}x$ and $\varphi_{\tau}$ is the Levy density of $\{g^{\tau}_t\}_{t\geq 0}$. Theorem 30.1 in \cite{Sato} also shows that
\begin{align}\label{glambdachar}
\E^{\Q}[e^{i\upsilon \tilde{g}^{\lambda}_t}]
=\exp\left(
	t\int_0^{\infty}\left(
		e^{x\log\left(\psi_{\lambda}(\upsilon)\right)}-1\right)
		\varphi_{\tau}(x)dx
	\right),
\end{align}
for $\upsilon\in \R$, where $\varphi_{\tau}$ is the Levy density of $\{g_t^{\tau}\}_{t\geq 0}$ and $\psi_{\lambda}(\upsilon)
:=\E\left[\exp\{i\upsilon g^{\lambda}_1\}\right]$. Note, in particular, that, since $|\psi_{\lambda}(\upsilon)|\leq 1$, \begin{align*}
Re(\log(\psi_{\lambda}(\upsilon)))
=\log(|\psi_{\lambda}(\upsilon))|)\leq 0,
\end{align*}
so the RHS of (\ref{glambdachar}) is indeed well defined. An application of Ito's lemma shows that, for $T\geq t\geq 0$,
\begin{equation}
\vec{X}_T
= e^{-\vec{\Theta}(T-t)}\vec{X}_t
	+ \int_t^Te^{-\vec{\Theta}(T-u)}d\vec{g}_u,
\end{equation}
and the process $\{\vec{X}_t\}_{t\geq 0}$ is then an OU process driven by the Levy process $\{\tilde{\vec{g}}_t\}_{t\geq 0}$. Note that, if $\rho$ is positive, then $r_t$ and $\lambda_t$ are positive for every $t\geq 0$. 

Throughout the rest of the paper, $C_0(\R^2)$ denotes the set of continuous functions on $\R^2$ vanishing at infinity, and $C_K^2(\R^2)$ and $C_0^2(\R^2)$ denote the set of twice continuously differentiable functions on $\R^2$ with compact support and vanishing at infinity respectively.

\begin{theorem}\label{BigTheorem}
The infinitesimal generator $\mathcal{A}$ of the Markov process $\{\vec{X}_t\}_{t\geq 0}$ is the smallest closed extension in the Banach space $C_0(\R^2)$ of the operator $\mathcal{A}_0$ defined, for $f\in C_K^2(\R^2)$ and $\vec{x}\in\R^2_+$, by
\begin{align}\label{Generator}
\mathcal{A}_0f(\vec{x})
 = - \vec{x}^*\vec{\Theta}\nabla f(\vec{x})
   + \int_{\R^2_+}\left(f(\vec{x}+\vec{y})-f(\vec{x})\right)
   		\varphi(d\vec{y})
\end{align}
where $\varphi$ is the Levy measure of $\{\tilde{\vec{g}}_t\}_{t\geq 0}$. Furthermore, $\varphi$ satisfies, for $f\in C^2_0(\R^2)$,
\begin{align}\label{Generator_g}
\int_{\R^2_+}\left(f(\vec{x}+\vec{y})-f(\vec{x})\right)\varphi(\vec{y})d\vec{y}
=\int_0^{\infty}f(r,\rho r)\varphi_r(r)dr
	+\int_0^{\infty}f(0,\lambda)\varphi_{\lambda}(\lambda)d\lambda,
\end{align}
where $\varphi_r$ and $\varphi_{\lambda}$ denote the Levy densities of $\{g^r_t\}_{t\geq 0}$ and $\{g^{\lambda}_t\}_{t\geq 0}$ respectively. In particular, the Levy measure of $\{g^{\lambda}_t\}_{t\geq 0}$ is absolutely continuous and $\varphi_{\lambda}$ is defined by 
\begin{align}\label{GammasubGammaLevy}
\varphi_{\lambda}(y)
&=\int_{0}^{\infty}p_{x}(y)
		\varphi_{\tau}(x)dx
 =\gamma_{\tau}\frac{e^{-{c_{\lambda}}y}}{y}\int_{0}^{\infty}
		\frac{(c_{\lambda}y)^{\gamma_{\lambda} x}}{\Gamma(\gamma_{\lambda} x)}
		\frac{e^{-c_{\tau} x}}{x}dx, \ y>0.
\end{align}
where $\Gamma$ denotes the Gamma function.
\end{theorem}
\begin{proof}
From \cite{Sato}, Theorem 31.5, $\mathcal{A}_0$ is the infinitesimal generator of the semigroup of contractions defined by $\{\vec{g}_t\}_{t\geq 0}$, except for the term $-\vec{x}^{\ast}\vec{\Theta}\nabla f(\vec{x})$. By \cite{Sato1984}, Theorem 3.1, its smallest closed extension in $C_0(\R^2)$ is the generator of the semigroup of contractions of $\{\vec{X}_t\}_{t\geq 0}$.\footnote{In fact, the process analyzed in \cite{Sato1984} is defined as the process defined by the trnasition semigroup associated to the the smallest close extension of $\mathcal{A}_0$. That such process coincide with $\{\vec{X}_t\}_{t\geq 0}$ is implied by the equivalence of the respective characteristic functions (equation (3.2) in \cite{Sato1984} is equivalent, after a change of variable, to equation (\ref{CharExp}) below with $\alpha_1=\alpha_2=0$).} To show \ref{Generator_g}, observe that
\begin{align*}
\E^{\Q}\left[e^{i\vec{u}^*\vec{g}_t}\right]
& = \E^{\Q}\left[e^{ig^r_t(u_r+\rho u_{\lambda})}\right]
	\E^{\Q}\left[e^{ig^{\lambda}_tu_{\lambda}}\right]\\
& = \exp\left\lbrace
	it\left(
	\int_0^{\infty}\left(e^{i(u_rr+ u_{\lambda}\rho r)}-1\right)\varphi_r(y)dr
	+\int_0^{\infty}\left(e^{iu_{\lambda}y}-1\right)
		\varphi_{\lambda}(\lambda)d\lambda
	\right)\right\rbrace,
\end{align*}
where $\vec{u}=(u_r,u_{\lambda})^*\in \R^2$ and since $\{g^r_t\}_{t\geq 0}$ and $\{g^{\lambda}_t\}_{t\geq 0}$ are independent. Furthermore, from (\ref{glambdachar}), the characteristic function of the process $\{\tilde{g}^{\lambda}_t\}_{t\geq 0}$ satisfies, for every $\upsilon\in \R$, $t\geq 0$,
\begin{align*}
\E^{\Q}\left[e^{i\upsilon\tilde{g}^{\lambda}_t}\right]
&=\exp\left(
	t\int_0^{\infty}\left(
		e^{x\log\left(\psi_{\lambda}(\upsilon)\right)}-1\right)
		\varphi_{\tau}(x)dx
	\right)\\
&=\exp\left(
	t\int_0^{\infty}\left(\psi_{\lambda}(\upsilon)^x-1\right)
	\varphi_{\tau}(x)dx\right)\\
&=\exp\left(
	t\int_0^{\infty}\left(\int_{0}^{\infty}
		e^{i\upsilon y}p_{x}(y)dy-1\right)
		\varphi_{\tau}(x)dx
	\right)\\
&=\exp\left(
	t\iint_{\R_+^2}
		\left(e^{i\upsilon y}-1\right)p_{x}(y)
		\varphi_{\tau}(x)dydx
	\right)\\
&=\exp\left(
	t\int_0^{\infty}
		\left(e^{i\upsilon y}-1\right)
		\left(\int_{0}^{\infty}p_{x}(y)
		\varphi_{\tau}(x)dx\right)dy
	\right),
\end{align*}
where, as before, $\psi_{\lambda}(\upsilon)
:=\E\left[\exp\{i\upsilon g^{\lambda}_1\}\right]$, and $p_{x}$ is the density of a gamma distribution with parameters $c_{\lambda}$ and $\gamma_{\lambda}x$. This shows \ref{GammasubGammaLevy}. The use of Fubini's Theorem in the last step of the above calculation can be justified as follows. First, by Tonelli's Theorem and Theorem 1.6 in \cite{BatirGammaInequalities},
\begin{align*}
\iint_{\R_+^2}
		|e^{i\upsilon y}-1|
		p_{x}(y)	\varphi_{\tau}(x)dxdy
& = \gamma_{\tau}\int_0^{\infty}
		|e^{i\upsilon y}-1|
		\frac{e^{-c_{\lambda}y}}{y}
		\int_{0}^{\infty}
		\frac{(c_{\lambda}y)^{\gamma_{\lambda} x}}	
		{\Gamma(\gamma_{\lambda} x)}
		\frac{e^{-c_{\tau} x}}{x}dxdy\\
& = \gamma_{\lambda}\gamma_{\tau}\int_0^{\infty}
		|e^{i\upsilon y}-1|
		\frac{e^{-c_{\lambda}y}}{y}
		\int_{0}^{\infty}
		\frac{(c_{\lambda}y)^{\gamma_{\lambda} x}}	
		{\Gamma(\gamma_{\lambda} x+1)}
		e^{-c_{\tau} x}dxdy\\
& \leq \gamma_{\lambda}\gamma_{\tau}\int_0^{\infty}
		|e^{i\upsilon y}-1|
		\frac{e^{-c_{\lambda}y}}{y}
		\left[\frac{1}{\hat{\Gamma}}
		\int_0^{\frac{1}{\gamma_{\lambda}}}
		(c_{\lambda}y)^{\gamma_{\lambda} x}
			e^{-c_{\tau} x}	dx
		\right.\\
	& \ \ \ \ \ \ \ \ \ \ \ \ \ \qquad \qquad 
		+\left.\int_{\tfrac{1}{\gamma_{\lambda}}}^{\infty}
		\frac{(c_{\lambda}y)^{\gamma_{\lambda} x}
			e^{-c_{\tau} x}}	
		{(\gamma_{\lambda} x)^{\gamma_{\lambda} x}
			e^{-\gamma_{\lambda} x}
			\sqrt{2\pi(\gamma_{\lambda} x+a)}}
		dx\right]dy
\end{align*}
where $\hat{\Gamma}=\min\{\Gamma(u), \ 1\leq u\leq 2\}$ and $a$ is a positive constant. Next, note that $|e^{i\upsilon y}-1|=O(y)$ as $y\rightarrow 0$ and that, on one hand,
\begin{align*}
\int_0^{\frac{1}{\gamma_{\lambda}}}
		(c_{\lambda}y)^{\gamma_{\lambda} x}
			e^{-c_{\tau} x}dx
& = \frac{1-e^{-\left(c_{\tau}
				-\gamma_{\lambda}\log(c_{\lambda}y)\right)
				/\gamma_{\lambda}}}	
		{\left(c_{\tau}
				-\gamma_{\lambda}\log(c_{\lambda}y)\right)}\\
& =  \frac{1-e^{-c_{\tau}/\gamma_{\lambda}}
				(c_{\lambda}y)}	
		{c_{\tau}
				-\gamma_{\lambda}\log(c_{\lambda}y)}
\end{align*}
and the last quantity is $O(1/\log(y))$ as $y\rightarrow 0$ and $O(y/\log(y))$ as $y\rightarrow\infty$. On the other hand,
\begin{align*}
\int_{\tfrac{1}{\gamma_{\lambda}}}^{\infty}
		\frac{(c_{\lambda}y)^{\gamma_{\lambda}x}}
		{(\gamma_{\lambda}x)^{\gamma_{\lambda}x}
			e^{-\gamma_{\lambda} x}}
		\frac{e^{-c_{\tau}x}}{\sqrt{\gamma_{\lambda} x+a}}dx
& = 	\int_{\tfrac{1}{\gamma_{\lambda}}}^{\infty}
		\frac{e^{-x(c_{\tau}
			+\gamma_{\lambda}[1-\log(c_{\lambda}y)]}}
		{(\gamma_{\lambda}x)^{\gamma_{\lambda}x}
		\sqrt{\gamma_{\lambda} x+a}}dx\\
& \leq 	\int_{\tfrac{1}{\gamma_{\lambda}}}^{\infty}
		\frac{f(y)^{x}}
		{(\gamma_{\lambda}x)^{\gamma_{\lambda}x}
		\sqrt{\gamma_{\lambda} x+a}}dx\\
& = 	\int_{\tfrac{1}{\gamma_{\lambda}}}^{\infty}
		\frac{e^{-x[\gamma_{\lambda}\log(\gamma_{\lambda}x)
			-\log(f(y)]}}
		{\sqrt{\gamma_{\lambda} x+a}}dx,
\end{align*}
where $f(y):= \exp\{\gamma_{\lambda}[\log(c_{\lambda}y)-1]-c_{\tau}\}\vee 1$. Observe now that
\begin{align*}
\int_{\tfrac{1}{\gamma_{\lambda}}}^{\infty}
		\frac{e^{-x[\gamma_{\lambda}\log(\gamma_{\lambda}x)
			-\log(f(y)]}}
		{\sqrt{\gamma_{\lambda} x+a}}dx 
&= \int_{\tfrac{1}{\gamma_{\lambda}}}
	 	^{\tfrac{(f(y)e)^{1/\gamma_{\lambda}}}
	 		{\gamma_{\lambda}}}
		\frac{e^{-x[\gamma_{\lambda}\log(\gamma_{\lambda}x)
			-\log(f(y)]}}
		{\sqrt{\gamma_{\lambda} x+a}}dx\\
	& \ \ 
	+ \int
		_{\tfrac{(f(y)e)^{1/\gamma_{\lambda}}}
			{\gamma_{\lambda}}}
	 	^{\infty}
		\frac{e^{-x[\gamma_{\lambda}\log(\gamma_{\lambda}x)
			-\log(f(y)]}}
		{\sqrt{\gamma_{\lambda} x+a}}dx \\
& \leq \frac{f(y)^{1/\gamma_{\lambda}}}{\sqrt{1+a}}
			\frac{(f(y)e)^{1/\gamma_{\lambda}}}
			{\gamma_{\lambda}}
			+\int_0^{\infty}\frac{e^{-x}}
				{\sqrt{\gamma_{\lambda} x+a}}dx,
\end{align*}
and the last term is $O(y^2)$ as $y\rightarrow \infty$ and $O(1)$ as $y\rightarrow 0$, and the result follows.
\end{proof}

The following result provides some basic features of the process $\{\tilde{g}^{\lambda}_t\}_{t\geq 0}$.

\begin{proposition}\label{GammaSubProp} The process $\{\tilde{g}^{\lambda}_t\}_{t\geq 0}$ almost surely has countably many and dense jumping times on $(0,\infty)$, and is of finite variation.
\end{proposition}
\begin{proof}
By Theorem 21.3 in \cite{Sato}, the first assertion follows on showing that $\varphi_{\lambda}$ is not integrable. Using again the relationship $u\Gamma(u)=\Gamma(u+1)$ for $u>0$, we obtain,
\begin{equation}\label{Ineq}
\begin{aligned}
\int_0^{\infty}
		\frac{(c_{\lambda}y)^{\gamma_{\lambda}x}}
		{\Gamma(\gamma_{\lambda}x)}
		\frac{e^{-c_{\tau}x}}{x}dx
& = \gamma_{\lambda}\int_0^{\infty}
		\frac{e^{-x[c_{\tau}
			-\gamma_{\lambda}\log(c_{\lambda}y)]}}
		{\Gamma(\gamma_{\lambda}x+1)}dx
  \geq \gamma_{\lambda}\int_0^{\tfrac{1}{\gamma_{\lambda}}}
		e^{-x[c_{\tau}
			-\gamma_{\lambda}\log(c_{\lambda}y)]}dx\\
& = \gamma_{\lambda}
		\frac{1-e^{-(c_{\tau}
			-\gamma_{\lambda}\log(c_{\lambda}y))/
				{\gamma_{\lambda}}}}
			{c_{\tau}
			-\gamma_{\lambda}\log(c_{\lambda}y)}
 = \frac{1-e^{-c_{\tau}/\gamma_{\lambda}}
				(c_{\lambda}y)}	
		{c_{\tau}
				-\gamma_{\lambda}\log(c_{\lambda}y)},
\end{aligned}
\end{equation}
which is $O(1/\log(y))$ as $y\rightarrow 0$. Hence,
\begin{align*}
\int_0^{\infty}\varphi_{\lambda}(y)dy
\geq
\gamma_{\lambda}
\int_0^{\infty}\frac{e^{-{c_{\tau}}y}}{y}
		\frac{1-e^{-\varepsilon[c_{\lambda}
			-\gamma_{\lambda}\log(c_{\lambda}y)]}}
			{c_{\lambda}	-\gamma_{\lambda}\log(c_{\lambda}y)}dy
= +\infty.
\end{align*}
To show that $\{\tilde{g}^{\lambda}_t\}_{t\geq 0}$ is of finite variation, note that 
\begin{align*}
\gamma_{\tau}\int_0^{1}\sqrt{y}\frac{e^{-c y}}{y}dy<\infty.
\end{align*}
The assertion then follows from Theorems 30.1 and 21.9 in \cite{Sato}.
\end{proof}

Next, we show that the distribution $\mu_{t_1,t_2}$ of the increment $\vec{X}_{t_2}-\vec{X}_{t_1}$ of the process $\{\vec{X}_t\}_{t\geq 0}$ is self decomposable for every $0\leq t_1\leq t_2$, i.e. there are sequences of (1) random variables $\{\vec{X}^{t_1,t_2}_n\}_{n \in\N }$, (2) invertible linear transformations $\{\vec{A}^{t_1,t_2}_n\}_{n\in \N}$ and, (3), vectors $\{\vec{a}^{t_1,t_2}_n\}_{n\in\N}$ such that
\begin{align*}
\vec{A}_n\sum_{i=1}^n\vec{X}_i^{t_1,t_2}-\vec{a}_n
& \xrightarrow[]{d} \vec{X}_{t_2}-\vec{X}_{t_1} \quad
\P(\|\vec{A}_n\vec{X}_n^{t_1,t_2}\|\geq \varepsilon)
  \rightarrow 0,
\end{align*}
for every $\varepsilon >0$, where $\xrightarrow[]{d}$ denotes convergence in distribution.
\begin{theorem}
The distribution $\mu_{t_1,t_2}$ of the increment $\vec{X}_{t_2}-\vec{X}_{t_1}$ of the vector process $\{\vec{X}_t\}_{t\geq 0}$ is self decomposable for every $0\leq t_1\leq t_2$.
\end{theorem}
\begin{proof}
According to a result by Urbanik (see \cite{Urbanik} and also \cite{Sato1984}), it is sufficient to show that
\begin{align}\label{LogInt}
\int_{\|\vec{x}\|\geq 1}\log(\|\vec{x}\|)\varphi(\vec{x})d\vec{x}
& = \int_{\frac{1}{1+\rho}}^{\infty}\log(x(1+\rho))\varphi_r(x)dx
	+\int_{1}^{\infty}\log(y)\varphi_{\lambda}(y)dy<\infty,
\end{align}
which can be obtained following a similar argument as in the proof of Theorem \ref{BigTheorem}.
\end{proof}
We conclude this subsection with the following remark.
\begin{remark} From (\ref{CDXSpreadFormula}), assuming only one semiannual payment and if $\{\vec{X}_t\}_{t\geq 0}$ is specified by (\ref{GDSRDI}), the credit spread is, in first order approximation, an affine linear function of the hazard rate:
\begin{align*}
\kappa(t,\lambda_t,T_0,T_M)
&=2\delta\left[
	K_1(t,T_0)e^{\lambda_t
		\left(\frac{1-e^{-\theta_{\lambda}/2}}
		{\theta_\lambda}\right)}-1\right]\approx \delta K_1(t,T_0)\lambda_t+K_2(t,T_0),
\end{align*}
where $K_1(t,T_0)$ and $K_2(t,T_0)$ are constants that do not depend on $\lambda_t$ nor $r_t$.\end{remark}


\subsection{Characteristic Exponent, Zero Coupon Bond Prices and Valuation PIDE}
In this section we compute the joint characteristic exponent of $\{\vec{X}_t\}_{t\geq 0}$ and its integrated process. To simplify notation, we henceforth drop the tilde and denote by $\{g^{\lambda}_t\}_{t\geq 0}$ the double gamma process $\{\tilde{g}^{\lambda}_t\}_{t\geq 0}$. Note that the integrated OU process $\{Y^r_t\}_{t\geq 0}$ associated to $\{r_t\}_{t\geq 0}$ satisfies, for $T\geq t\geq 0$,
\begin{align*}
Y^r_T-Y^r_t
& = \int_t^Tr_sds 
  = \int_t^Tr_te^{-\theta_r(s-t)}ds
 	+\int_t^T\int_t^se^{-\theta_r(s-u)}dg^r_uds \\
& = r_t\left(\frac{1-e^{-\theta_r(T-t)}}{\theta_r}\right)
 	+\int_t^T\frac{1-e^{-\theta_r(T-u)}}{\theta_r}dg^r_u
\end{align*}
Similarly, the integrated OU process $Y^{\lambda}_t$ associated to $\lambda_t$ satisfies
\begin{align*}
Y^{\lambda}_T-Y^{\lambda}_t
& = \lambda_t\left(\frac{1-e^{-\theta_{\lambda}(T-t)}}
	{\theta_{\lambda}}\right)
 	+\int_t^T\frac{1-e^{-\theta_{\lambda}(T-u)}}
 	{\theta_{\lambda}}d(\rho g^r_u+g^{\lambda}_u)
\end{align*}
Therefore, for every $\alpha_1,\alpha_2,\alpha_3,\alpha_4 \in\R$, we have
\begin{equation}\label{CharFun}
\begin{aligned}
& \alpha_1 (Y^r_T-Y^r_t)+\alpha_2 (Y^{\lambda}_T-Y^{\lambda}_t)
+\alpha_3r_T+\alpha_4\lambda_T\\
& \ 
  = \alpha_1	r_t\left(\frac{1-e^{-\theta_r(T-t)}}{\theta_r}\right)
 	+\alpha_2\lambda_t\left(\frac{1-e^{-\theta_{\lambda}(T-t)}}
	 	{\theta_{\lambda}}\right)
 	+\alpha_3r_te^{-\theta_r(T-t)}
 	+\alpha_4\lambda_te^{-\theta_{\lambda}(T-t)}\\
	& \ \ 
	+\int_t^T\left[
	\alpha_1\left(\frac{1-e^{-\theta_r(T-u)}}
	{\theta_r}\right)
	+\alpha_2\rho
		\left(\frac{1-e^{-\theta_{\lambda}(T-u)}}
	 	{\theta_{\lambda}}\right)
	+\alpha_3\left(e^{-\theta_r(T-u)}\right)
	+\alpha_4\rho
		\left(e^{-\theta_{\lambda}(T-u)}\right)
		\right]dg^r_u\\
	& \ \ 
	+\int_t^T\left[
		\alpha_2
		\left(\frac{1-e^{-\theta_{\lambda}(T-u)}}
	 	{\theta_{\lambda}}\right)
	+\alpha_4\left(e^{-\theta_{\lambda}(T-u)}\right)
	\right]dg^{\lambda}_u.
\end{aligned}
\end{equation}
Next, set
\begin{align*}
\xi_r(t,r,\alpha_1,\alpha_3)
& := \alpha_1r
	\left(\frac{1-e^{-\theta_rt}}{\theta_r}\right)
	+\alpha_3re^{-\theta_rt}\\
\xi_{\lambda}(t,\lambda,\alpha_2,\alpha_4)
& := \alpha_2\lambda\left(\frac{1-e^{-\theta_{\lambda}t}}
 	{\theta_{\lambda}}\right)
 	+\alpha_4\lambda e^{-\theta_{\lambda}t}\\
\psi^r_u(t,\alpha_1,\alpha_2,\alpha_3,\alpha_4)
& := \alpha_1\left(\frac{1-e^{-\theta_r(t-u)}}
	{\theta_r}\right)
	+\alpha_2\rho
		\left(\frac{1-e^{-\theta_{\lambda}(t-u)}}
	 	{\theta_{\lambda}}\right)\\
 	& \ \ \
	+\alpha_3\left(e^{-\theta_r(t-u)}\right)
	+\alpha_4\rho e^{-\theta_{\lambda}(t-u)}\\
\psi^{\lambda}_u(t ,\alpha_2,\alpha_4)
& := \alpha_2
		\left(\frac{1-e^{-\theta_{\lambda}(t-u)}}
	 	{\theta_{\lambda}}\right)
	+\alpha_4\left(e^{-\theta_{\lambda}(t-u)}\right).
\end{align*}
Since, for every $0<t\leq u\leq T$,
\begin{align*}
0& < |\psi^r_u(T,\alpha_1,\alpha_2,\alpha_3,\alpha_4)|
	\leq |\alpha_1|+\rho|\alpha_2|+|\alpha_3|+\rho|\alpha_4| \\
0& <|\psi^{\lambda}_u(T,\alpha_2,\alpha_4)|\leq+\rho(|\alpha_2|+|\alpha_4|),
\end{align*}
the stochastic integrals 
\begin{align*}
\Psi^r_t(T,\alpha_1,\alpha_2,\alpha_3,\alpha_4)
	 & := \int_t^T\psi^r_u(T,\alpha_1,\alpha_2,\alpha_3,\alpha_4)
	 		 dg^r_u 
	 	\stackrel{d}{=}
	 	  \int_0^{T-t}\psi^r_{T-u}
	 	  	(T,\alpha_1,\alpha_2,\alpha_3,\alpha_4)dg^r_u  \\
\Psi^{\lambda}_t(T,\alpha_2,\alpha_4)
	 & := \int_t^T\psi^r_u(T,\alpha_1,\alpha_2,\alpha_3,\alpha_4) 			dg^{\lambda}_u
	 	\stackrel{d}{=}
	 	  \int_0^{T-t}\psi^r_{T-u}(T,\alpha_1,\alpha_2,\alpha_3,\alpha_4) 	dg^{\lambda}_u
\end{align*}
appearing in (\ref{CharFun}) are well defined for $0\leq t \leq u\leq T$ and their jumps correspond to jumps in the processes $\{g^r_t\}_{T\geq t\geq 0}$ and $\{g^{\lambda}_t\}_{T\geq t\geq 0}$. By Proposition IX.5.3 in \cite{JacodShiryaev}, the third characteristics $\{\nu^r_t\}_{t\geq 0}$ and $\{\nu^{\lambda}_t\}_{t\geq 0}$ of the semimartingales $\{\Psi^r_{T-t\wedge T}\}_{t\geq 0}$ and $\{\Psi^{\lambda}_{T-t\wedge T}\}_{t\geq 0}$ satisfy
\begin{align*}
\nu^r_t(G) & = \int_{0}^t\int_0^{\infty}\indic{G}(T-u,\psi_{T-u}^r(T,\alpha_1,\alpha_2,\alpha_3,\alpha_4)y)\varphi_r(y)dydu\\
\nu^{\lambda}_t(G) & = \int_{0}^t\int_0^{\infty}\indic{G}(T-u,\psi_{T-u}^{\lambda}(T,\alpha_2,\alpha_4)y)\varphi_{\lambda}(y)dydu,
\end{align*}
for every $0\leq t\leq T$, and where $G$ is a Borel subset of $[0,t]\times (0,\infty)$. Since $\{\nu^r_t\}_{t\geq 0}$ and $\{\nu^{\lambda}_t\}_{t\geq 0}$ are deterministic, $\{\Psi^r_{T-t\wedge T}\}_{t\geq 0}$ and $\{\Psi^{\lambda}_{T-t\wedge T}\}_{t\geq 0}$ have independent increments. By Theorem II.4.15 in \cite{JacodShiryaev}, the process $\{\Psi^r_{t\wedge T}\}_{t\geq 0}$ satisfies, for every $\upsilon \in\R$,
\begin{align*}
& \E^{\Q}\left[\left.\exp\left\lbrace\int_t^T
	i\upsilon \psi^r_u(T,\alpha_1,\alpha_2,\alpha_3,\alpha_4)dg^r_u
	\right\rbrace\right\rvert\F_t\right]\\
& \ \ =
    \exp\left\lbrace\int_t^T
	\int_0^{\infty}\left(
	e^{i\upsilon\psi^r_u(T,\alpha_1,\alpha_2,\alpha_3,
	\alpha_4)y}-1\right)
	\varphi_r(y)dydu\right\rbrace\\
& \ \ = \exp\left(-\gamma_{r}\int_t^T
	\log\left(1-\frac{i\upsilon\psi^r_u(T,\alpha_1,\alpha_2,\alpha_3,\alpha_4)}
	{c_{r}}\right)du\right).
\end{align*}
Note that the right hand side is defined for $\upsilon$ purely imaginary, as long as $i\upsilon\psi_u^r(T,\alpha_1,\alpha_2,\alpha_3,\alpha_4)<c_r$ for every $T\geq u\geq t$, and so, by Theorem 25.17 in \cite{Sato}, the above equivalences hold also if $\upsilon = i$ and the coefficients $\alpha_1$, $\alpha_2$, $\alpha_3$, $\alpha_4$ are nonnegative, as, in this case, $\psi^r_u(T,\alpha_1,\alpha_2,\alpha_3,\alpha_4)$ is nonnegative for $t\leq u\leq T$.

Similarly, we obtain for every $\upsilon \in \R$,
\begin{align*}
&\E^{\Q}\left[\left.\exp
\left\lbrace\int_t^Ti\upsilon\psi^{\lambda}_u(T,\alpha_2,\alpha_4)
	dg^{\lambda}_u\right\rbrace\right\rvert \F_t\right]\\
& \ \ = \exp\left\lbrace\int_t^{T}
	\int_0^{\infty}\left(
	e^{i\upsilon\psi^{\lambda}_u(T,\alpha_2,\alpha_4)y}-1\right)
	\varphi_{\lambda}(y)dydu\right\rbrace\\
& \ \ = \exp\left\lbrace-\gamma_{\tau}\int_t^{T}
	\log\left(1+\frac{\gamma_{\lambda}}{c_{\tau}}\log
	\left(1-\frac{i}{c_{\lambda}}
	\upsilon\psi^{\lambda}_u(T,\alpha_2,\alpha_4)
	\right)\right)
	du\right\rbrace
\end{align*}
with the equivalences holding also if $\upsilon = i$ and $\alpha_2$ and $\alpha_4$ are nonnegative.
Hence, for $T\geq t\geq 0$, $\upsilon \in \{1,i\}$ and coefficients $\alpha_1$, $\alpha_2$, $\alpha_3$, $\alpha_4$ that are real if $\upsilon =1$ and nonnegative if $\upsilon =i$,
\begin{equation}\label{CharExp}
\begin{aligned}
& \log\left(\E^{\Q}\left[e^{i\upsilon\alpha_1 (Y^r_T-Y^r_t)
+i\upsilon\alpha_2 (Y^{\lambda}_T-Y^{\lambda}_t)+i\upsilon\alpha_3r_T+i\upsilon\alpha_4\lambda_T}
|\F_t\right]\right)\\
& \ \ \
  = \int_t^{T}
	\int_0^{\infty}\left(
	e^{i\upsilon\psi^r_u(T,\alpha_1,\alpha_2,\alpha_3,
	\alpha_4)y}-1\right)
	\varphi_r(y)dydu\\
& \ \ \ \ \ \ \  
  + \int_t^{T}
	\int_0^{\infty}\left(
	e^{i\upsilon\psi^{\lambda}_u(T,\alpha_2,\alpha_4)y}-1\right)
	\varphi_{\lambda}(y)dydu \\
& \ \ \ \ \ \ \
  + i\upsilon\xi_r(T-t,r_t,\alpha_1,\alpha_3)
	+i\upsilon\xi_{\lambda}(T-t,\lambda_t,\alpha_2,\alpha_4).
\end{aligned}
\end{equation}
This implies immediately that the risk neutral price $P(t,T)$ at time $t$ of a zero coupon bond with maturity $T$ and no default risk is given by
\begin{align*}
P(t,T)
& = \exp\left\lbrace
	-r_t\left(\frac{1-e^{-\theta_r(T-t)}}
	{\theta_r}\right)
	+\int_t^{T}
	\int_0^{\infty}\left(e^{-\frac{1-e^{-\theta_r(T-u)}}
	{\theta_r}y}-1\right)
	\varphi_r(y)dydu\right\rbrace.
\end{align*}
\subsubsection{Density, Summary Statistics and Stationary Distribution}\label{Stats}
\begin{figure*}
        \centering
        \begin{subfigure}[b]{0.450\textwidth}
            \centering
            \includegraphics[width=\textwidth]{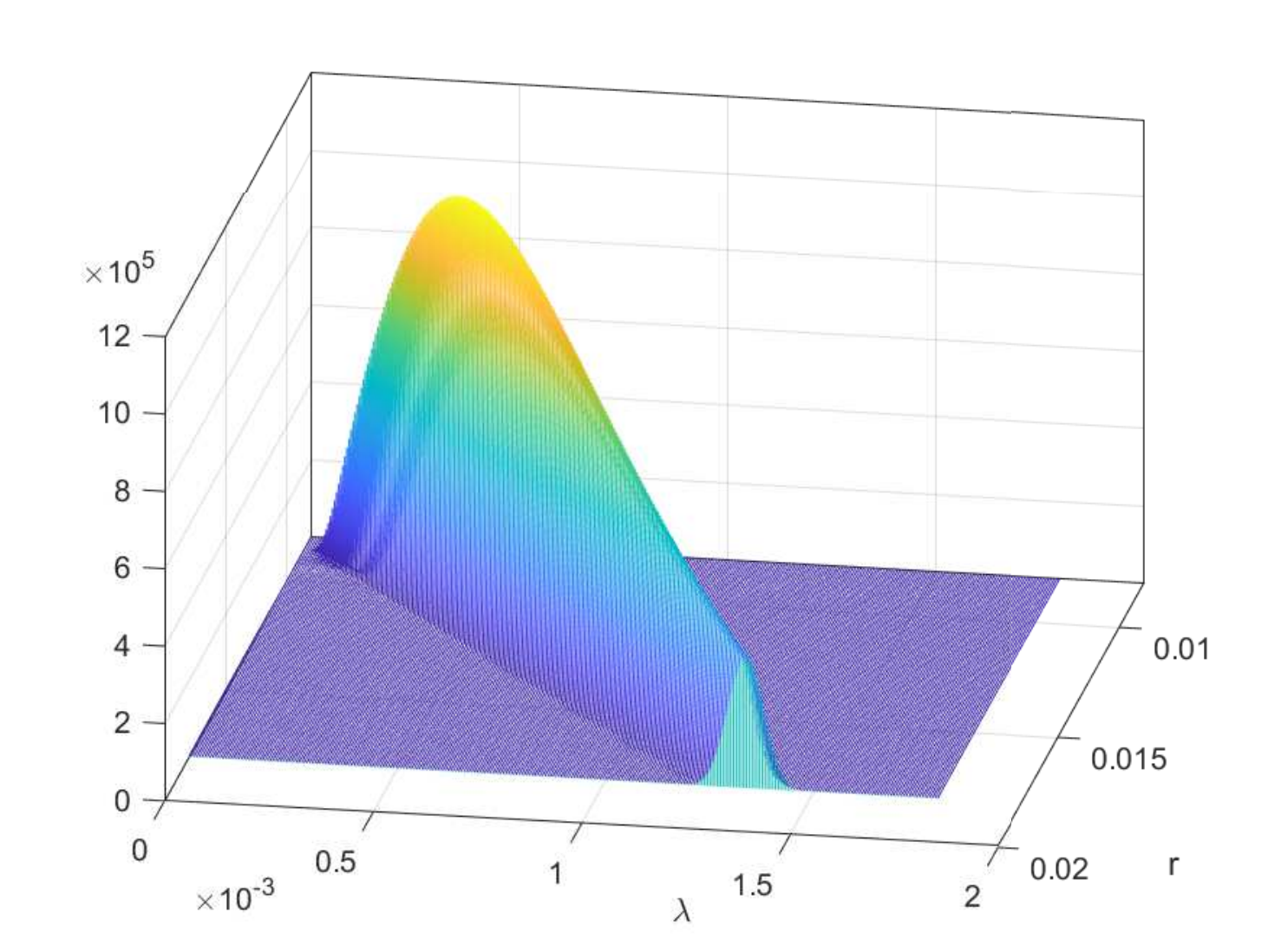}
            \caption[]%
            {{Bivariate density of $\vec{X}_t$ for $t=1$ year.}}    
        \end{subfigure}
        \hfill
        \begin{subfigure}[b]{0.450\textwidth}  
            \centering 
            \includegraphics[width=\textwidth]{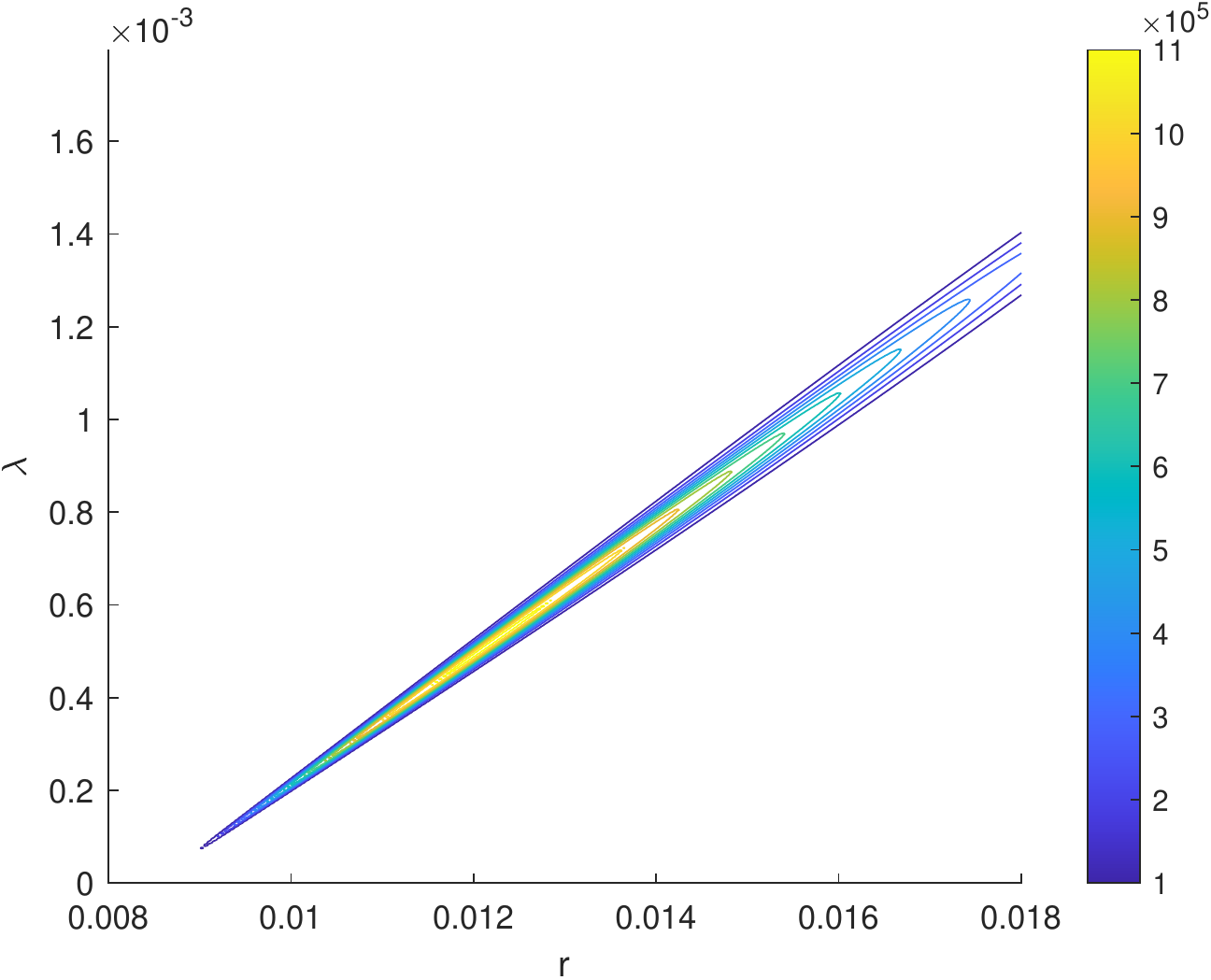}
            \caption[]%
            {{Bivariate density contour of $\vec{X}_t$ for $t=1$ year.}}  
        \end{subfigure} 
        \caption{}
        \label{BivDensities}
        \centering
        \begin{subfigure}[b]{0.450\textwidth}
            \centering
            \includegraphics[width=\textwidth]{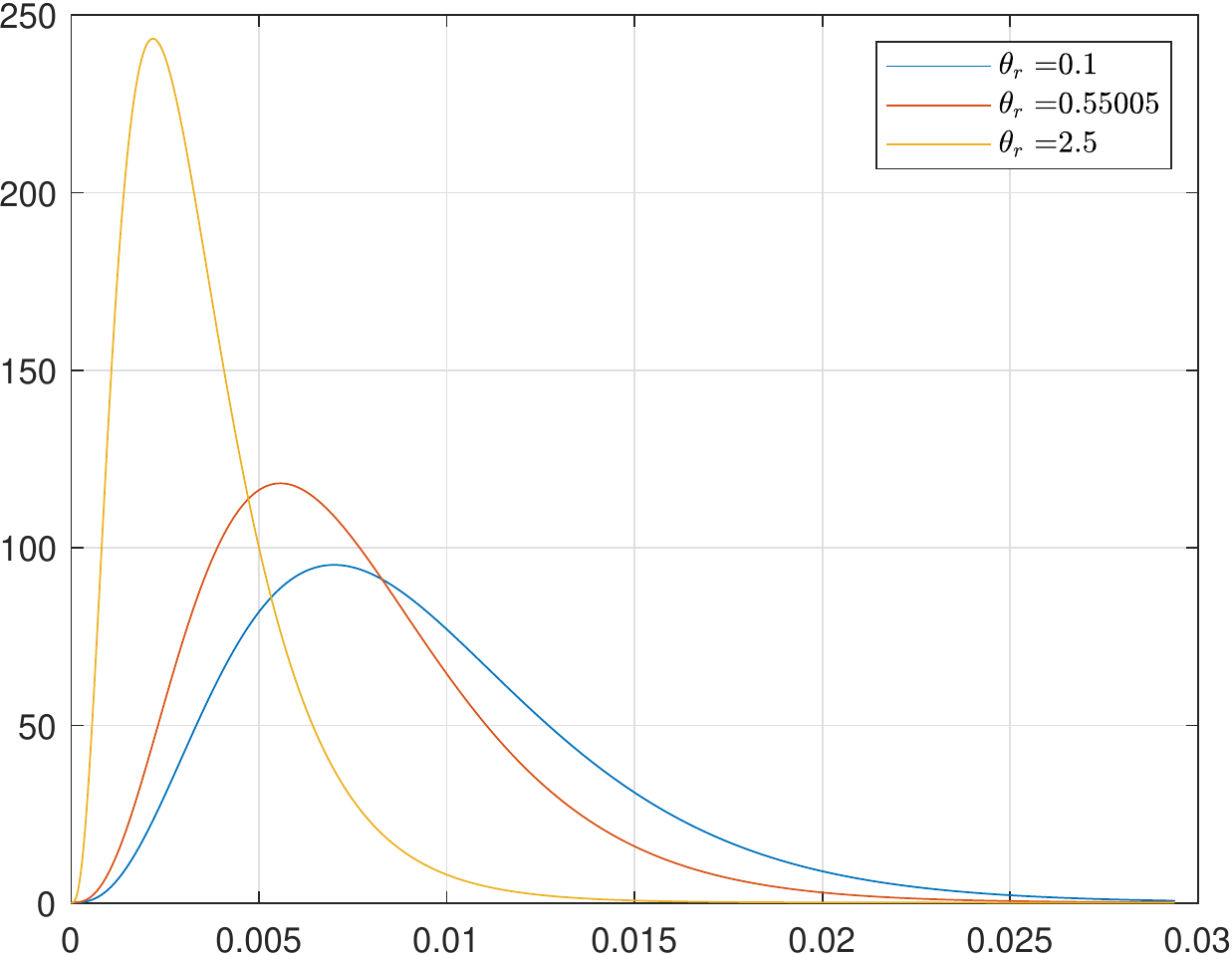}
            \caption[]%
            {{Marginal densities for the short rate for $\theta_r\in\{0.16,1,5\}$ and $t=1$ year.}}    
        \end{subfigure}
        \hfill
        \begin{subfigure}[b]{0.450\textwidth}  
            \centering 
            \includegraphics[width=\textwidth]{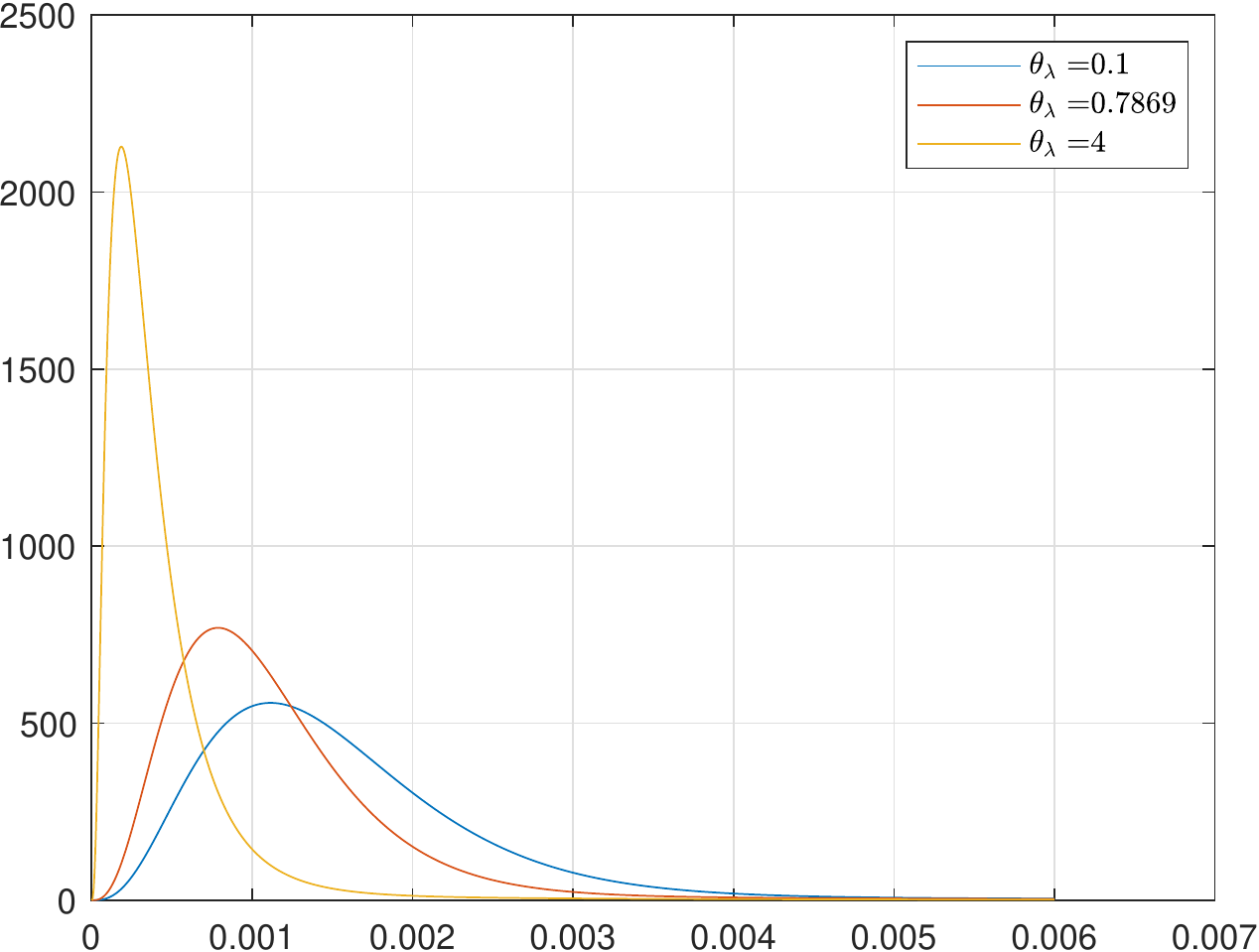}
            \caption[]%
            {{Default intensity marginal for $\theta_{\lambda}\in\{0.5,2.06,4\}$ and $t=1$ year.}}  
        \end{subfigure} 
        \caption{}
        \label{MarginalDensities}
        \begin{subfigure}[b]{0.450\textwidth}
            \centering
            \includegraphics[width=\textwidth]{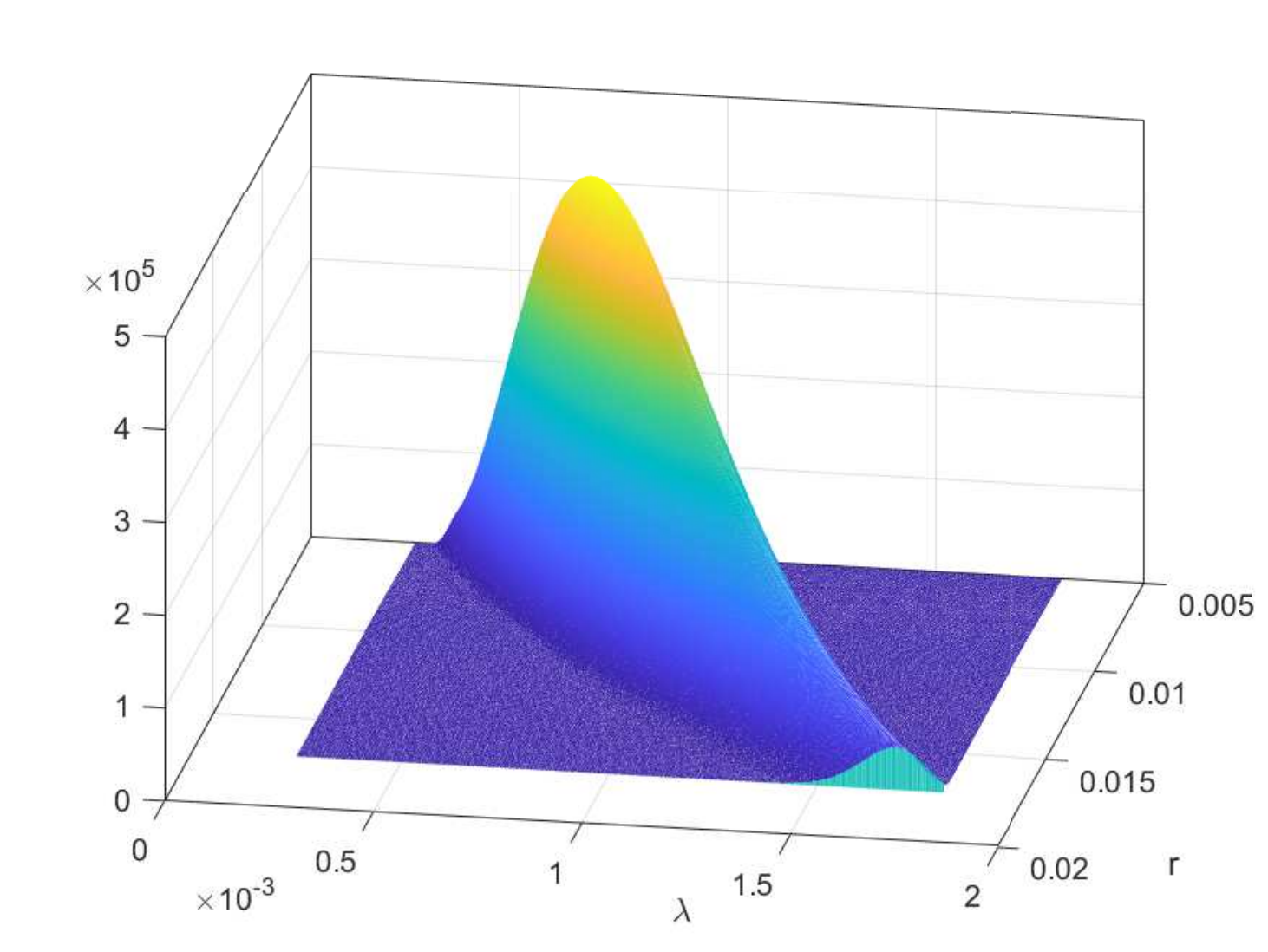}
            \caption[]%
            {{Stationary density of $\{\vec{X}_t\}_{t\geq 0}$.}}    
        \end{subfigure}
        \hfill
        \begin{subfigure}[b]{0.450\textwidth}  
            \centering 
            \includegraphics[width=\textwidth]{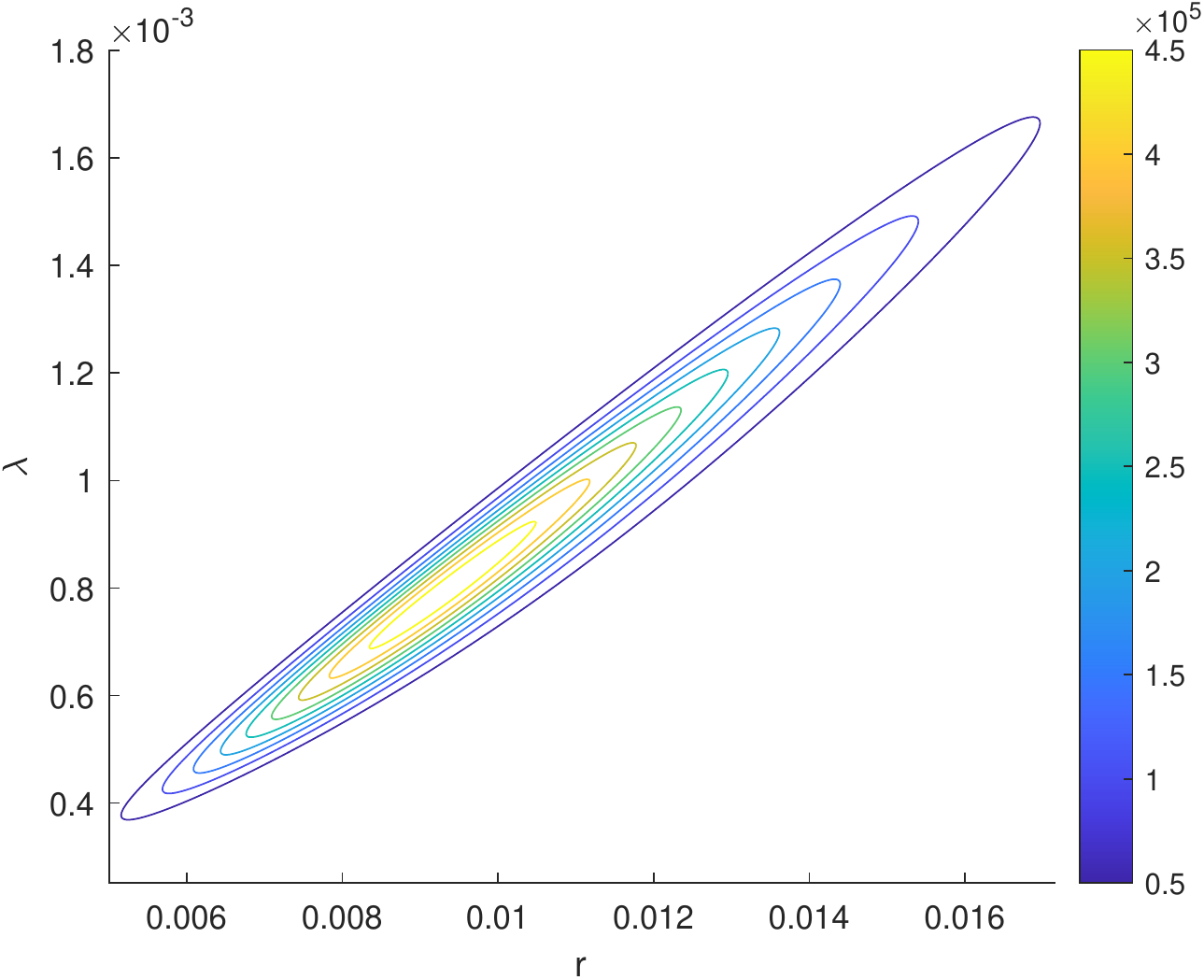}
            \caption[]%
            {{Stationary density contour of $\{\vec{X}_t\}_{t\geq 0}$.}}  
        \end{subfigure}
        \caption{}
        \label{BivDensitiesStationary}
\end{figure*}

To gain further insights into the dynamics of the process $\{\vec{X}_t\}_{t\geq 0}$, we will now derive its density $f_{r,\lambda,t}$ at time $t\geq 0$ given its initial state $(r_0,\lambda_0)^*$. Denoting by $\phi_t$ the (joint) Fourier transform, we have, for $\vec{\alpha}=(\alpha_1,\alpha_2)^*\in\R^2$,
\begin{equation*}
\begin{aligned}
\log\left(\phi_t(\vec{\alpha})\right)
& = \log\left(\iint_{\R^2}e^{-2\pi i(\alpha_1 r+\alpha_2\lambda)}
	f_{r,\lambda,t}(r,\lambda)drd\lambda\right)
  = \log\left(\E^{\Q}\left[\left.
	e^{-2\pi i(\alpha_1 r_t+\alpha_2 \lambda_t)}\right]\right\rvert \F_0\right)
	\notag\\
& = -2\pi i\alpha_1 r_0e^{-\theta_r t}
	-2\pi i\alpha_2 \lambda_0 e^{-\theta_{\lambda}t}\notag\\
	& \ \ \
	-\gamma_{r}\int_0^t\log\left(1+\frac{2\pi i}{c_{r}}
		\left(\alpha_1 e^{-\theta_r(t-u)}
		+\rho\alpha_2 e^{-\theta_{\lambda}(t-u)}\right)
		\right)du\notag\\
	& \ \ \
	-\gamma_{\tau}\int_0^t
		\log\left(1+\frac{\gamma_{\lambda}}{c_{\tau}}
		\log\left(1+\frac{2\pi i}{c_{\lambda}}
		\alpha_2 e^{-\theta_{\lambda}(t-u)}\right)
		\right)du
\end{aligned}
\end{equation*}
By Fourier inversion and a change of variable, the joint density of $\vec{X}_t$ for $t\geq 0$ is then given by
\begin{align}\label{Density}
f_{r,\lambda,t}(r,\lambda)
= \frac{1}{4\pi^2}\iint_{\R^2}e^{
	i(\alpha_1 r+\alpha_2\lambda)}
	\phi_t\left(\frac{\alpha_1}{2\pi},
	\frac{\alpha_2}{2\pi}\right)d\alpha_1 d\alpha_2
\end{align}
As shown in \cite{Hurd}, this double integral can be computed using a two dimensional fast Fourier transform. Specifically, we set $N=2^{13}$, $B=10^6$, $\eta=\frac{2B}{N}$, $\lambda=\frac{2\pi}{N\eta}=\frac{\pi}{B}$, $b=\frac{N\lambda}{2}=\frac{\pi}{\eta}$, and approximate (\ref{Density}) by a double sum over the grid in the frequency domain
\begin{equation*}
F = \left\lbrace\alpha_{k}
	= \left(\alpha_{k_1},\alpha_{k_2}\right):
				k = (k_1,k_2)\in
				\left\lbrace 0,1,...,N-1\right\rbrace^2
				\right\rbrace, \
\alpha_{k_i}=-B+k_i\eta, \ i=1,2
\end{equation*}
with corresponding grid in the space domain given by
\begin{equation*}
S = \left\lbrace x_{\ell}
	= \left(x_{\ell_1},x_{\ell_2}\right):
				\ell = (\ell_1,\ell_2)\in
				\left\lbrace 0,1,...,N-1\right\rbrace^2
				\right\rbrace, \
x_{\ell_i}=-b+\ell_i\eta, \ i=1,2.
\end{equation*}
Thus, we have the approximation
\begin{align*}
f_{r,\lambda,t}(r,\lambda) 
&\approx
\frac{\eta^2}{4\pi^2}\sum_{k_1,k_2=0}^{N-1}
e^{i\alpha_kx_{\ell}'}
	\phi_t\left(\frac{\alpha_{k_1}}{2\pi},
	\frac{\alpha_{k_2}}{2\pi}\right)\\
&=(-1)^{\ell_1+\ell_2}\left(\frac{\eta N}{2\pi}\right)^2
\frac{1}{N^2}\sum_{k_1,k_2=0}^{N-1}e^{2\pi ik\ell'/N}
(-1)^{k_1+k_2}\phi_t\left(\frac{\alpha_{k_1}}{2\pi},
	\frac{\alpha_{k_2}}{2\pi}\right),
\end{align*}
where the last double sum can be computed for instance in Matlab with the command $\tt{ifft2}$. Figures (\ref{BivDensities}) and (\ref{MarginalDensities}) show the bivariate density and the marginals of the vector $\{\vec{X}_t\}_{t\geq 0}$ for $t=1$ year, and for 
\begin{align*}
& r_0 = 0.0146, \ \theta_r=0.5500\,\ c_r = 400.0005,
\ \gamma_r = 3.9475,\ \rho = 0.1548;\\
& \lambda_0 = 0,\ 
	\theta_{\lambda}=3.3533,\ c_{\lambda}=4.3178, \
	\gamma_{\lambda} = 6.0617,\
	c_{\tau} = 3.5298,\ \gamma_{\tau} = 190.0001,
\end{align*}
which are the parameters obtained from calibration to CDX options traded on 2 January 2020 with maturity in 2 months (see section \ref{Experiments}).

\vspace{3mm}

The characteristic function allows one to compute summary statistics for $\{r_t\}_{t\geq 0}$ and $\{\lambda_t\}_{t\geq 0}$:
\begin{align*}
\E^{\Q}[r_t|\F_0]
 & = r_0e^{-\theta_rt}
 	+\frac{1}{i} \frac{\partial}{\partial \alpha}
 	\left[-\gamma_{r}\int_0^t\log\left(1-\frac{i\alpha e^{-\theta_r(t-u)}}{c_r}
	\right)du\right]_{\alpha=0}\\
&  = r_0e^{-\theta_r t}+\frac{\gamma_r}{c_r}
 	 \frac{1-e^{-\theta_rt}}{\theta_r},\\
\mathbb{V}^{\Q}[r_t|\F_0]
 & = -\frac{\partial^2}{\partial \alpha^2}
 	 	\left[-\gamma_{r}\int_0^t\log\left(1-\frac{i\alpha e^{-\theta_r(t-u)}}{c_r}
	\right)du\right]_{\alpha=0}
   = \frac{\gamma_r}{c_r^2}
 	 \frac{1-e^{-2\theta_rt}}{2\theta_r}\\
\E^{\Q}[\lambda_t|\F_0]
 & = \lambda_0e^{-\theta_{\lambda} t}
 	 +\left(\frac{\rho\gamma_r}{c_r}
 	 +\frac{\gamma_{\tau}}{c_{\tau}}\frac{\gamma_{\lambda}}{c_{\lambda}}\right)
 	 \frac{1-e^{-\theta_{\lambda}t}}{\theta_{\lambda}}\\
\mathbb{V}^{\Q}[\lambda_t|\F_0]
 & = \left[\frac{\rho^2\gamma_r}{c^2_r}
 	 +\frac{\gamma_{\tau}\gamma_{\lambda}}{c_{\tau}^2c_{\lambda}^2}
 	 \left(\gamma_{\lambda}+c_{\tau}\right)\right]\frac{1-e^{-2\theta_{\lambda}t}}{2\theta_{\lambda}}.
\end{align*}
We omit the long, but simple calculations, but we note that that higher values of the parameters $\theta_r$ and $\theta_{\lambda}$ imply smaller short rate and default intensity, whereas the smaller $\theta_r$ and $\theta_{\lambda}$, the higher the variance and expected value of $r_t$ and $\lambda_t$ respectively (see also figure \ref{MarginalDensities}). 

\vspace{3mm}

Finally, following again from
\begin{align*}
\int_{\|\vec{x}\|\geq 1}\log(\|\vec{x}\|)\varphi(\vec{x})d\vec{x}
& = \int_{\frac{1}{1+\rho}}^{\infty}\log(r(1+\rho))\varphi_r(r)dr
	+\int_{1}^{\infty}\log(\lambda)\varphi_{\lambda}(\lambda)d\lambda
	<\infty,
\end{align*}
and Theorem 4.1 in \cite{Sato1984}, $\{\vec{X}\}_{t\geq 0}$ admits a self decomposable stationary distribution whose Fourier transform is given by:
\begin{equation}\label{StationaryFourier}
\begin{aligned}
\phi(\alpha_1,\alpha_2)
& =  \exp\left\lbrace
	-\gamma_r\int_0^{\infty}\log\left(1+
		\frac{2\pi i}{c_r}\left(\alpha_1 e^{-\theta_rv}
		+\rho\alpha_2 e^{-\theta_{\lambda}v}\right)
		\right)dv\right.\\
& \ \ \
	\left.
	-\gamma_{\tau}\int_0^{\infty}\log\left(1+
		\frac{\gamma_{\lambda}}{c_{\tau}}
		\log\left(1+\frac{2\pi i}{c_{\lambda}}
		\alpha_2 e^{-\theta_{\lambda}v}\right)\right)dv
		\right\rbrace.
\end{aligned}
\end{equation}
Figure \ref{BivDensitiesStationary} shows the limiting density of the bivariate process $\{\vec{X}_t\}_{t\geq 0}$, obtained by Fourier inversion and computing the integrals in (\ref{StationaryFourier}) on the interval $[0,100]$.

%

\label{CDXPayoffsection}

\subsubsection{Linear PIDE for CDX Swaption prices}
By the First Fundamental Theorem of asset pricing, we know that the time $t$ price $u(t,\vec{X}_t)$ of a CDX swpation given $\F_t$ satisfies
\begin{equation*}
u(t,\vec{X}_t)
=\E^{\Q}\left[\left.e^{-\int_t^{T_0}r_udu}\pi(\vec{X}_{T_0})\right\rvert\F_t\right].
\end{equation*}
Therefore, the process $\{M_t\}_{t\geq 0}$ defined for every $t\in[0,T]$ by
\begin{equation*}
M_t
:=e^{-\int_0^{t}r_udu}u(t,\vec{X}_t)
\end{equation*}
is a local martingale. If $u\in C^{1,2}_K(\R\times \R^2)$, Ito's lemma for semimartingales gives for every $T_0\geq t\geq 0$,
\begin{align*}
M_t
& =M_0-\int_0^t
	e^{-\int_0^{s}r_udu}r_s
	u(s,\vec{X}_s)ds\\
& \ \ \ \
	+\int_0^te^{-\int_0^{s}r_udu}
	\left[u_t(s,\vec{X}_s)
	-\vec{X}_s^*\vec{\Theta}\nabla u(s,\vec{X}_s)
	\right]ds\\
& \ \ \ \
	+\iint_{(0,t]\times\R_+^2\setminus\{0\}}
	e^{-\int_0^{s}r_udu}\mathcal{D}_u^{s,\vec{X}_s}(\mathbf{y})
	N(ds,d\mathbf{y}),
\end{align*}
where $N$ is the random measure associated to the process $\{\vec{X}_t\}_{t\geq 0}$ and
$ \mathcal{D}_u^{t,\vec{x}}(\vec{y}):= u(t,\vec{x}+\vec{y})-u(t,\vec{x})$. 
Adding and subtracting the compensator of $N$, which can be computed again from Proposition IX.5.3 in \cite{JacodShiryaev}, and equating the drift to zero gives, for every $t\geq 0$,
\begin{align}\label{integraleq}
\int_0^te^{-\int_0^{s}r_udu}
	\left[u_t(s,\vec{X}_s)
	-\vec{X}_s^*\vec{\Theta}\nabla u(s,\vec{X}_s)
	+\int_{\R^2\setminus\{0\}}
	\mathcal{D}_u^{s,\vec{X}_s}(e^{-(t-s)\vec{\Theta}}\vec{y})
	\varphi(d\vec{y}) \right]ds = 0
\end{align}
where, we recall,
\begin{align*}
\int_{\R^2_+\setminus\{\vec{0}\}}
		\mathcal{D}_u^{t,\vec{x}}(\vec{y})\varphi(d\vec{y})
& = \int_0^{\infty}\left(
	u\left(t,\vec{x}+
		\begin{bmatrix}
			y_r \\
			\rho y_r
		\end{bmatrix}
		\right)
	-u(t,\vec{x})
	\right)
	\varphi_r(y_r)dy_r\\
	& \ \ \
	+\int_0^{\infty}\left(
	u\left(t,\vec{x}+
		\begin{bmatrix}
			0 \\
			y_{\lambda}
		\end{bmatrix}
	\right)
	-u(t,\vec{x})\right)
	\varphi_{\lambda}(y_{\lambda})dy_{\lambda}
\end{align*}
for all $\vec{x}\in\R^2_+\setminus\{\vec{0}\}$. Differentiating (\ref{integraleq}) and reversing time yields the following PIDE for the price function $u$:
\begin{equation}
\begin{cases}
u_t+ru-\mathcal{A}u(t,\cdot)(\vec{x})
	=0, \ t\in\R_+, \ \vec{x}\in\R^2_+\setminus\{\vec{0}\} \\
u(0,\vec{x})=\pi(T_0,\vec{x})\ \vec{x}\in\R^2_+\setminus\{\vec{0}\},
\end{cases}\label{PIDEiTraxxOpt}
\end{equation}
where $\mathcal{A}$, the infinitesimal generator of the Markov process $\{\vec{X}_t\}_{t\geq 0}$, is specified by (\ref{Generator}), based on Theorem \ref{BigTheorem} and the assumption that $u\in C^{1,2}_K(\R\times \R^2)$.

 
\section{Numerical Results}
We implemented a finite difference scheme for the valuation PIDE, whose construction is reported in the Appendix. The scheme was then tested taking as final condition the payoff of a forward start CDX swap, whose current value admits an integral representation. The proof is based on calculations that are similar to those performed in section \ref{CDXPayoffsection}. In particular, we obtain that, for every $t\in[0,T_0]$, $\ell=1,...,M$,
\begin{align*}
& \E^{\Q}\left[\left.
 	e^{-\int_{t}^{T_0}r_udu}
 	\E^{\Q}\left[e^{-\int_{T_0}^{T_\ell}(r_u+\lambda_u)du}
	|\F_{T_0}\right]\right\rvert\F_{t}\right] \\
& 
 = \E^{\Q}\left[\left.
 	e^{-(Y^r_{T_0}-Y^r_t)}
 	\E^{\Q}\left[e^{-(Y^r_{T_{\ell}}-Y^r_{T_0})-(Y^{\lambda}_{T_{\ell}}
 		+Y^{\lambda}_{T_0})}
	|\F_{T_0}\right]\right\rvert\F_{t}\right]\\
& 
 = \E^{\Q}\left[
 	e^{-(Y^{r}_{T_0}-Y^r_t)}
 	e^{-\xi_r(T_{\ell}-T_0,r_{T_0},1,0)
	-\xi_{\lambda}(T_{\ell}-T_0,\lambda_{T_0},1,0)}
	|\F_t\right]\\
	& \ \ \
 	\times e^{\int_{T_0}^{T_{\ell}}
	\int_0^{\infty}\left(
	e^{-\psi^r_u(T_{\ell},1,1,0,0)y}-1\right)
	\varphi_r(y)dydu}\\
	& \ \ \
	\times e^{\int_{T_0}^{T_{\ell}}
	\int_0^{\infty}\left(e^{-\psi^{\lambda}_u	
	(T_{\ell},1,0)y}-1\right)
	\varphi_{\lambda}(y)dydu}\\
& 
 = e^{-\xi_r(T_{0}-t,r_t,1,b_3)
	-\xi_{\lambda}(T_{0}-t,\lambda_t,0,b_4)}\\
	& \ \ \
	\times e^{\int_{t}^{T_0}
	\int_0^{\infty}\left(
	e^{-\psi^r_u(T_{0},1,0,b_3,b_4)y}-1\right)
	\varphi_r(y)dydu}\\
	& \ \ \
	\times e^{\int_{t}^{T_{0}}
	\int_0^{\infty}\left(e^{-\psi^{\lambda}_u	
	(T_{0},0,b_4)y}-1\right)
	\varphi_{\lambda}(y)dydu}\\
	& \ \ \
 	\times e^{\int_{T_0}^{T_{\ell}}
	\int_0^{\infty}\left(
	e^{-\psi^r_u(T_{\ell},1,1,0,0)y}-1\right)
	\varphi_r(y)dydu}\\
	& \ \ \
	\times
	e^{\int_{T_0}^{T_{\ell}}
	\int_0^{\infty}\left(e^{-\psi^{\lambda}_u
	(T_{\ell},1,0)y}-1\right)
	\varphi_{\lambda}(y)dydu},
\end{align*}
and, similarly, 
\begin{align*}
& \E^{\Q}\left[\left.
 	e^{-\int_t^{T_0}r_udu}
 	\E^{\Q}\left[e^{-\int_{T_0}^{T_{\ell-1}}(r_u+\lambda_u)du}
	P(T_{\ell-1},T_{\ell})|\F_{T_0}\right]
	\right\rvert\F_{t}\right]\\
& 
 = \E^{\Q}\left[\left.
 	e^{-(Y^r_{T_0}-Y^r_t)}
	\E^\Q\left[\left.e^{-(Y^r_{T_{\ell -1}}-Y^r_{T_0})
		-(Y^{\lambda}_{T_{\ell -1}}-Y^{\lambda}_{T_0})+r_{T_{\ell-1}}\zeta}
		\right\rvert\F_{T_0}\right]
	\right\rvert\F_{t}\right]\\
	& \ \ \
	\times e^{
	\int_{T_{\ell-1}}^{T_{\ell}}\int_0^{\infty}
	\left(e^{-\frac{1-e^{-\theta_r(T_{\ell}-u)}}
	{\theta_r}y}-1\right)\varphi_r(y)dydu}\\
&
 = \E^{\Q}\left[\left.
 	e^{-(Y^r_{T_0}-Y^r_t)}
	e^{-\xi_r(T_{\ell-1}-T_0,r_{T_0},1,\zeta)
	-\xi_{\lambda}(T_{\ell-1}-T_0,\lambda_{T_0},1,0)}
	\right\rvert\F_{t}\right]\\
	& \ \ \
	\times e^{ \int_{T_0}^{T_{\ell-1}}
	\int_0^{\infty}\left(
	e^{-\psi^r_u(T_{\ell-1},1,1,\zeta,0)y}-1\right)
	\varphi_r(y)dydu
	+\int_{T_0}^{T_{\ell-1}}
	\int_0^{\infty}\left(
	e^{-\psi^{\lambda}_u(T_{\ell-1},1,0)y}-1\right)
	\varphi_{\lambda}(y)dydu}\\
	& \ \ \
	\times e^{
	\int_{T_{\ell-1}}^{T_{\ell}}\int_0^{\infty}
	\left(e^{-\frac{1-e^{-\theta_r(T_{\ell}-u)}}
	{\theta_r}y}-1\right)\varphi_r(y)dydu}\\
&
 =  e^{-\xi_r(T_{0}-t,r_t,1,a_3)
	-\xi_{\lambda}(T_{0}-t,\lambda_t,0,a_4)}\\
	& \ \ \
	\times e^{
	\int_{t}^{T_0}
	\int_0^{\infty}\left(
	e^{-\psi^r_u(T_{0},1,0,a_3,a_4)y}-1\right)
	\varphi_r(y)dydu
	+\int_{t}^{T_{0}}
	\int_0^{\infty}\left(
	e^{-\psi^{\lambda}_u(T_{0},0,a_4)y}-1\right)
	\varphi_{\lambda}(y)dydu}.\\
	& \ \ \
	\times e^{\int_{T_0}^{T_{\ell-1}}
	\int_0^{\infty}\left(
	e^{-\psi^r_u(T_{\ell-1},1,1,\zeta,0)y}-1\right)
	\varphi_r(y)dydu
	+\int_{T_0}^{T_{\ell-1}}
	\int_0^{\infty}\left(
	e^{-\psi^{\lambda}_u(T_{\ell-1},1,0)y}-1\right)
	\varphi_{\lambda}(y)dydu}\\
	& \ \ \
	\times e^{
	\int_{T_{\ell-1}}^{T_{\ell}}\int_0^{\infty}
	\left(e^{-\frac{1-e^{-\theta_r(T_{\ell}-u)}}
	{\theta_r}y}-1\right)\varphi_r(y)dydu},
\end{align*}
where
\begin{align*}
&  \zeta:=
    \frac{1-e^{-\theta_r(T_{\ell}-T_{\ell-1})}}{\theta_r},\
  a_3 := \frac{1-e^{-\theta_r(T_{\ell-1}-T_{0})}}{\theta_r}
	+\zeta e^{-\theta_r (T_{\ell-1}-T_0)}, \\
&  a_4 := \frac{1-e^{-\theta_{\lambda}(T_{\ell-1}-T_{0})}}
	{\theta_{\lambda}}, \
  b_3:=
     \frac{1-e^{-\theta_r(T_{\ell}-T_{0})}}{\theta_r},\
  b_4 := \frac{1-e^{-\theta_{\lambda}(T_{\ell}-T_{0})}}
	{\theta_{\lambda}}.
\end{align*}

\begin{figure*}
        \centering
        \begin{subfigure}[b]{0.475\textwidth}
            \centering
            \includegraphics[width=\textwidth]
            {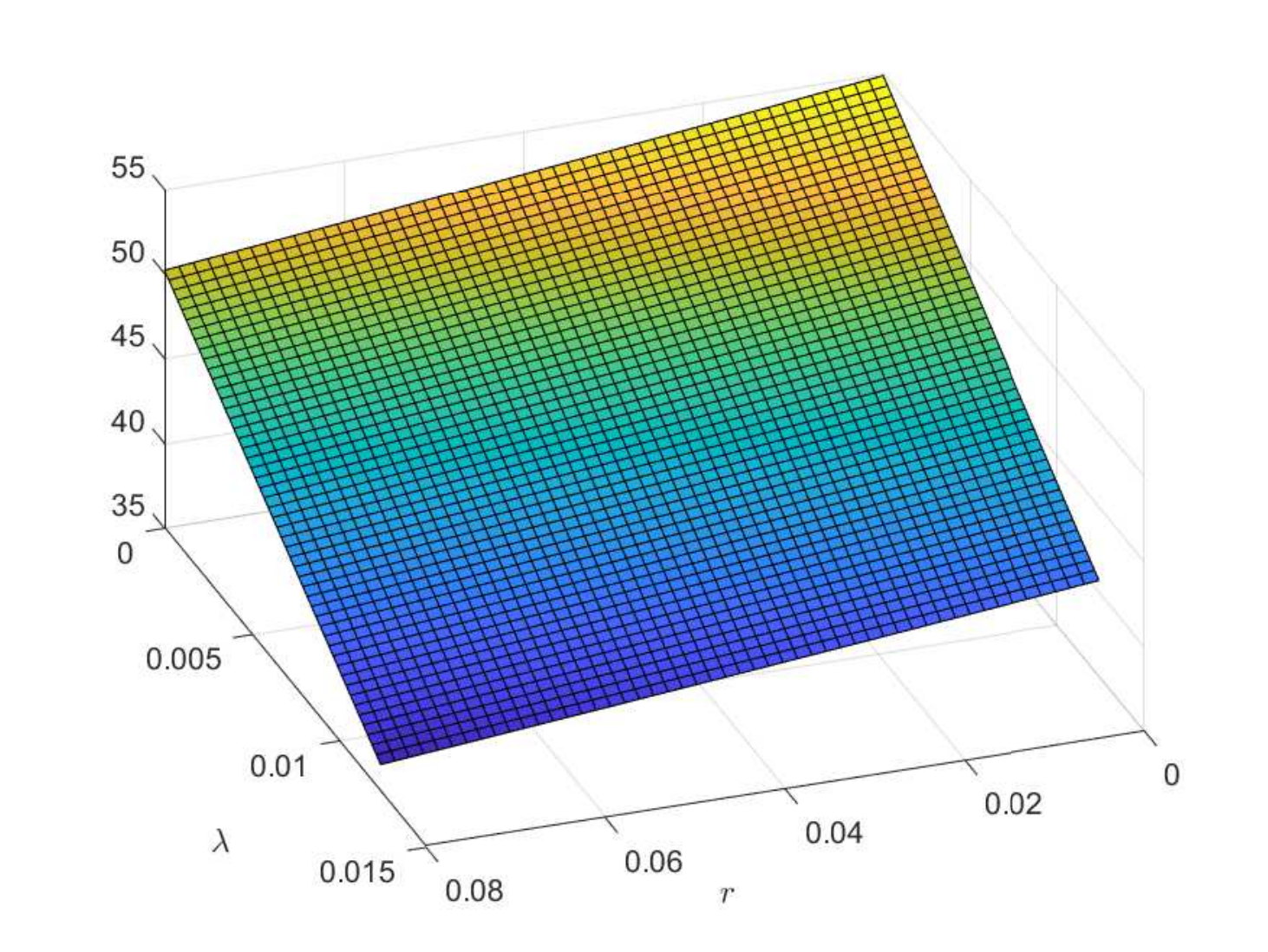}
            \caption[]%
            {{}}    
        \end{subfigure}
        \hfill
        \begin{subfigure}[b]{0.475\textwidth}  
            \centering 
            \includegraphics[width=\textwidth]
            {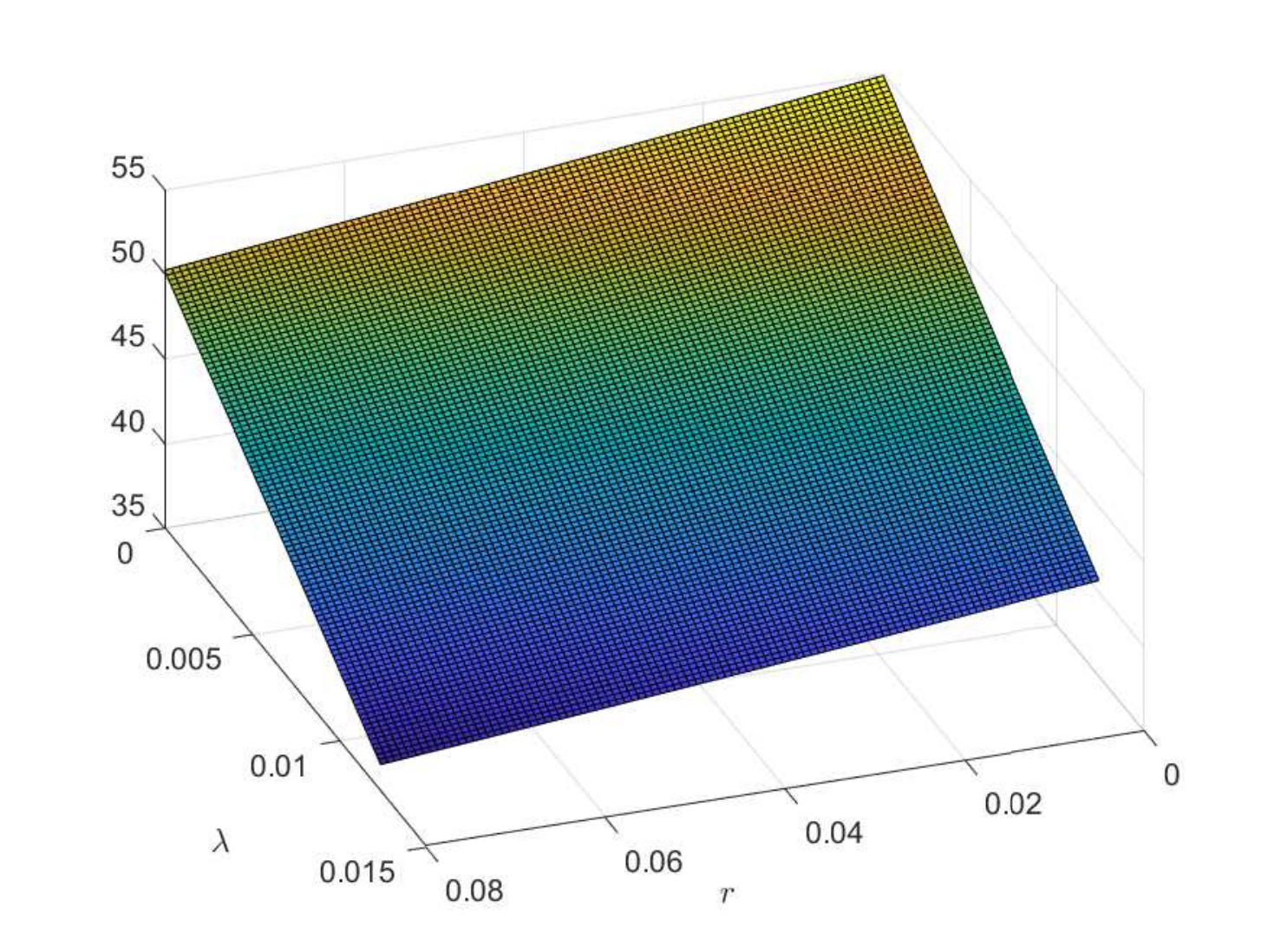}
            \caption[]%
            {{}}  
        \end{subfigure} 
        \caption{Numerical price surface (in bps) for a forward-start CDX, assuming $N=50$ (a) and $N=100$ (b), and for $M=100$ and $N_{sim}=100$.}
        \label{CDXNumerical}

        \centering
        \begin{subfigure}[b]{0.43\textwidth}
            \centering
            \includegraphics[width=\textwidth]
            {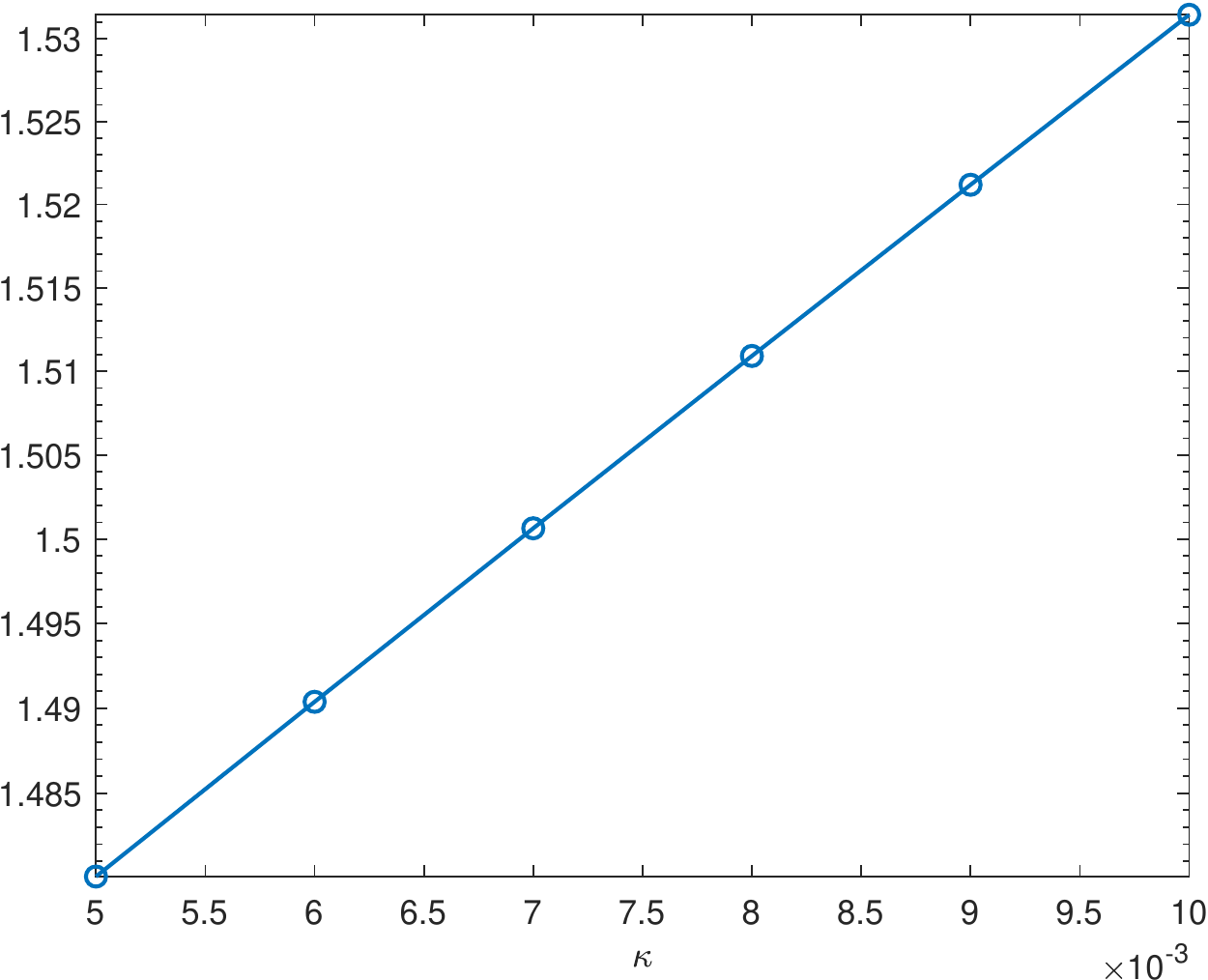}
            \caption[]%
            {{}}    
        \end{subfigure}
        \hfill
        \begin{subfigure}[b]{0.43\textwidth}  
            \centering 
            \includegraphics[width=\textwidth]
            {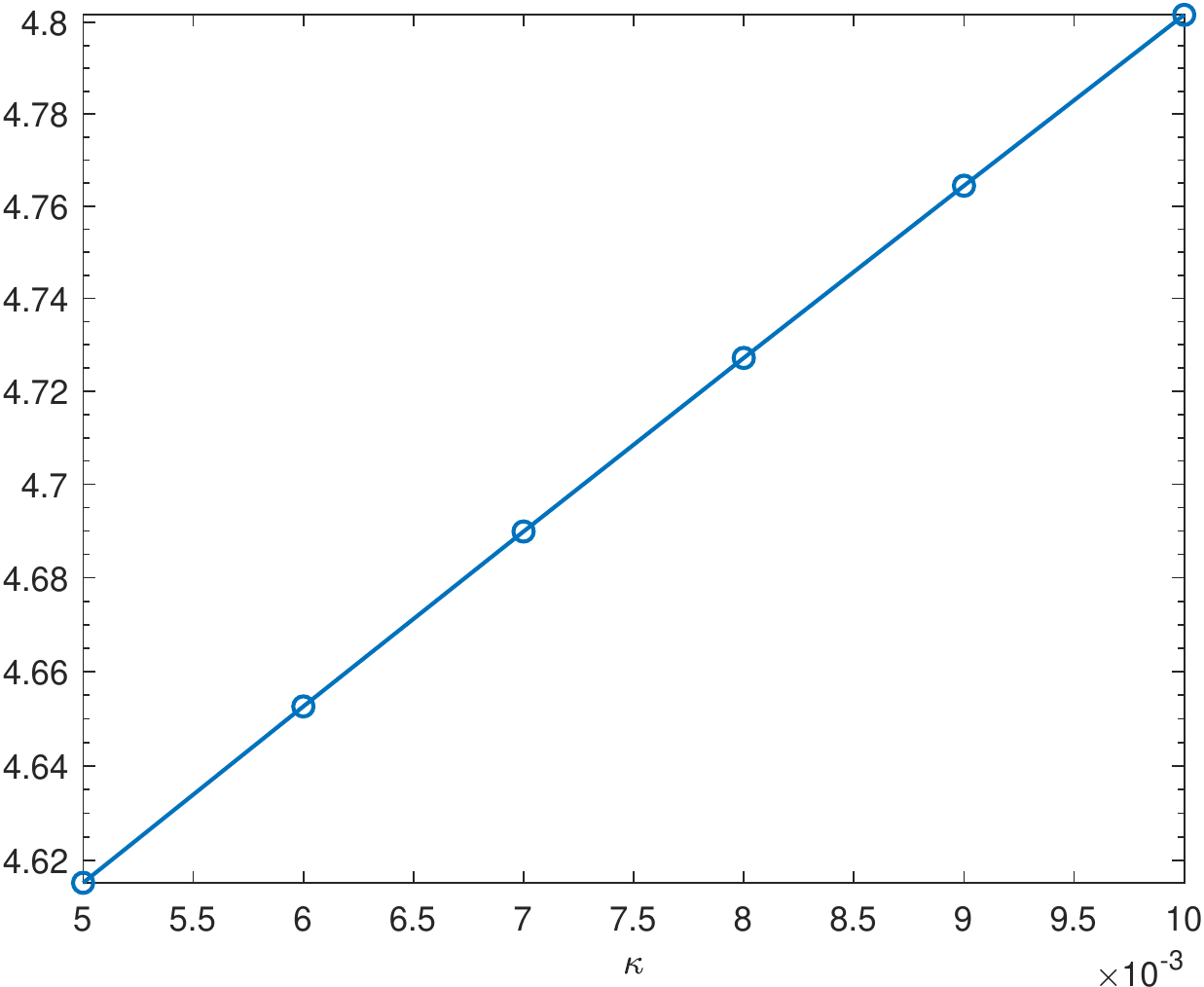}
            \caption[]%
            {{}}  
        \end{subfigure} 
        \caption{$\ell^{\infty}$ absolute error (a) and $\ell^{\infty}$ difference (in bps) with Montecarlo generated price surface (b) for strikes $\kappa=50:10:100$ bps.}
        \label{CDXerr}

        \centering
        \begin{subfigure}[b]{0.475\textwidth}
            \centering
            \includegraphics[width=\textwidth]{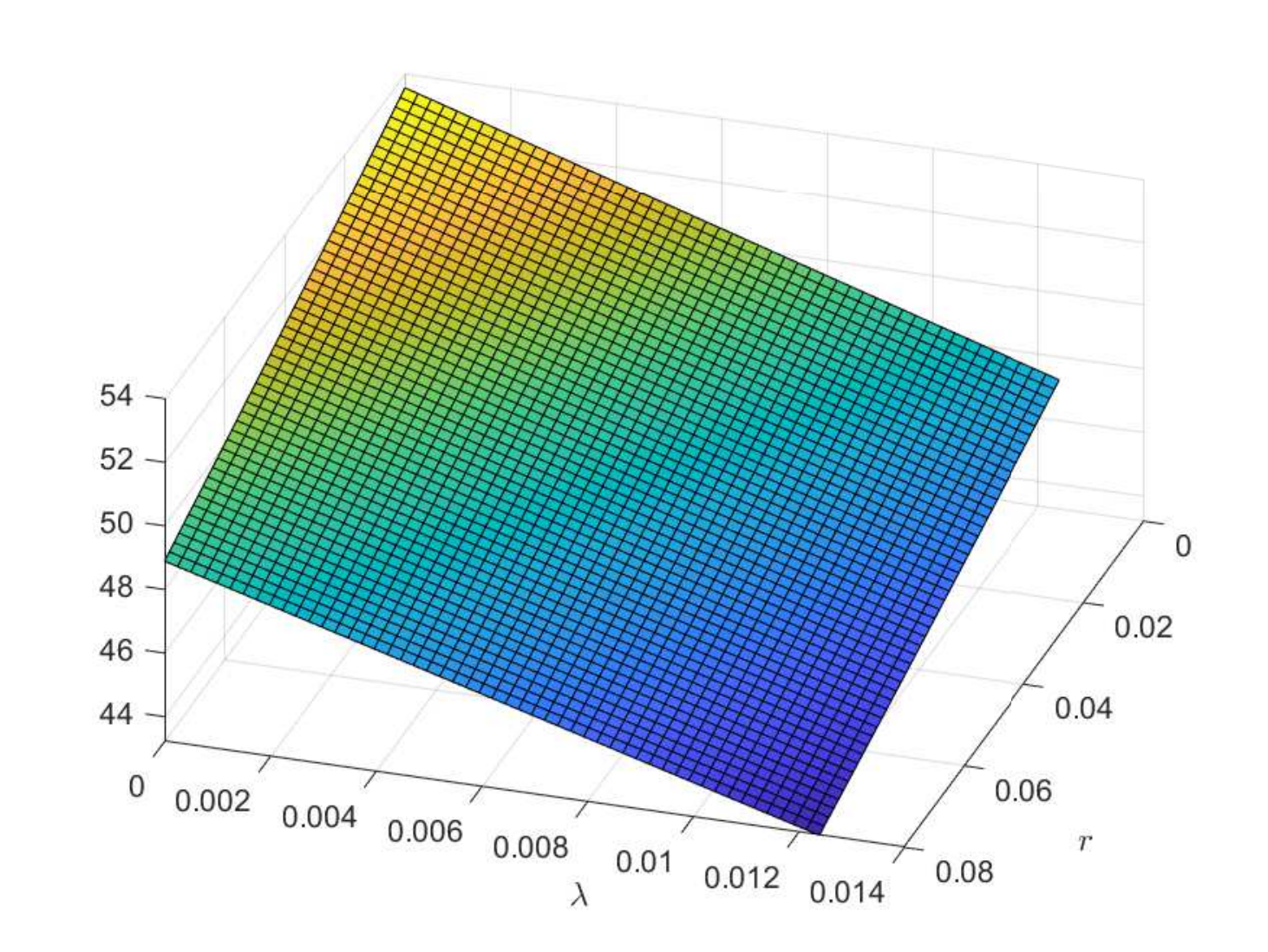}
            \caption[]%
            {{}}    
        \end{subfigure}
        \hfill
        \begin{subfigure}[b]{0.475\textwidth}  
            \centering 
            \includegraphics[width=\textwidth]{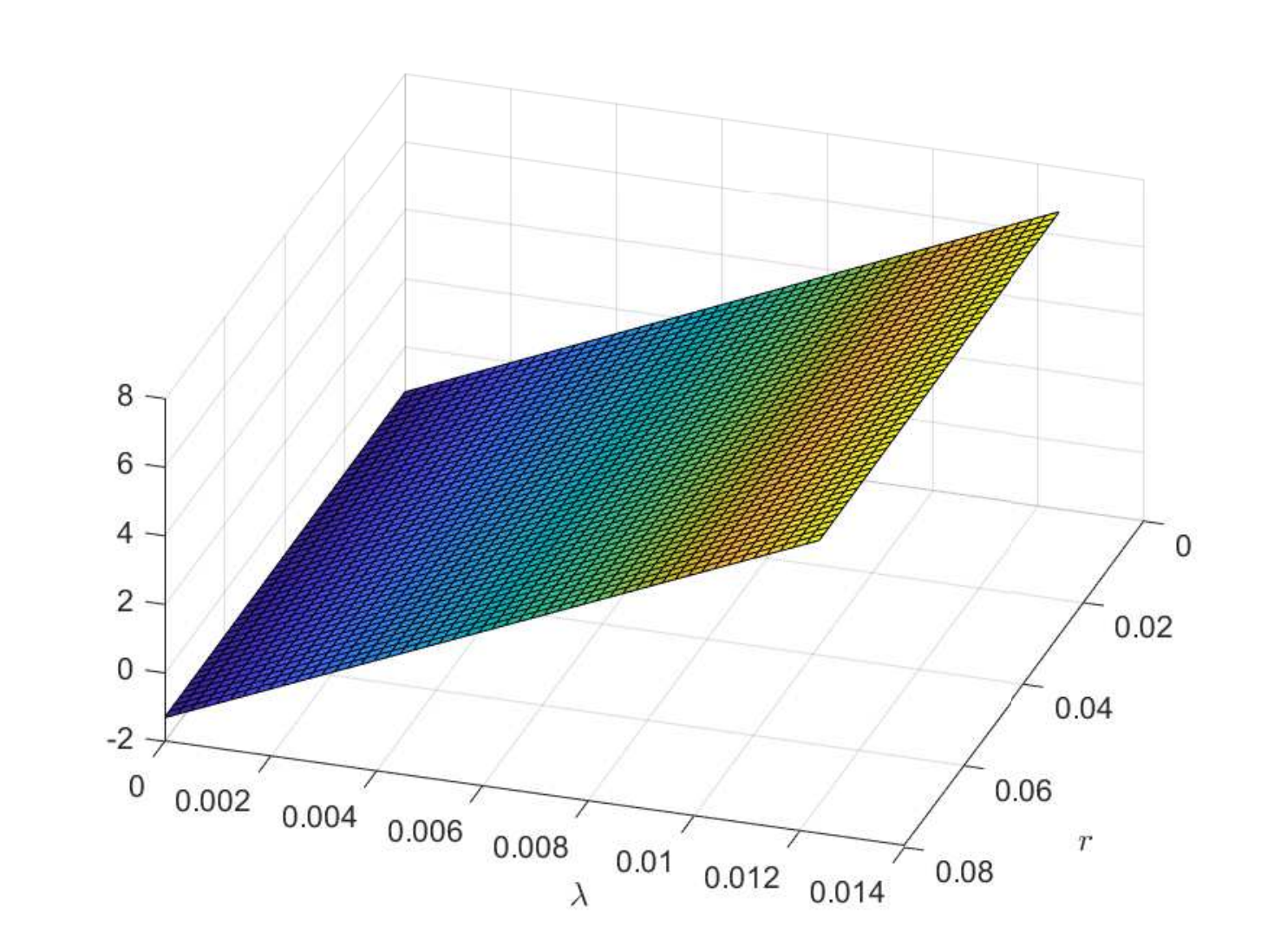}
            \caption[]%
            {{}}  
        \end{subfigure} 
        \caption{Price surface  (in bps) of a forward-start CDX including FEP on a $50\times50$ grid (a) and comparison with the numerical solution (excluding FEP) of (\ref{GDSRDIEqn}) for strike $\kappa=100$ bps.}
        \label{CDX}
\end{figure*}

We considered as before the following set of parameters,
\begin{align*}
& r_0 = 0.0146, \ \theta_r=0.5500\,\ c_r = 400.0005,
\ \gamma_r = 3.9475,\ \rho = 0.1548;\\
& \lambda_0 = 0,\ 
	\theta_{\lambda}=3.3533,\ c_{\lambda}=4.3178, \
	\gamma_{\lambda} = 6.0617,\
	c_{\tau} = 3.5298,\ \gamma_{\tau} = 190.0001,
\end{align*}
and we also assumed that the forward contract matures in 15 days, while the underlying asset is a $5$-year receiver swap with recovery rate of $0.4$, strike $\kappa=60$ bps and semiannual payments. Figure \ref{CDXNumerical} shows the price surface for the forward start CDX generated by solving (\ref{PIDEiTraxxOpt}) assuming $N=50$ (left) and $N=100$ (right), and with initial condition given by the payoff of the swap at maturity of the forward contract. The $\ell^{\infty}$ absolute error is plotted in figure \ref{CDXerr}(a) for strikes $\kappa=50:10:100$ bps. Note that even to compute the analytical solutions certain integrations were performed numerically. Figure \ref{CDXerr} (b) shows the $\ell^{\infty}$ absolute error of the solution computed via Montecarlo simulation. For both the PIDE and Montecarlo cases the error is relatively high, and although one could reduce the error for instance by more accurately computing the gamma time changed gamma L\'evy density (\ref{GammasubGammaLevy}) (at the price of higher computational costs), we observe that the error is less than or at least comparable with the bid-ask spread observed in the option market, which, in the period considered, is at least 2 bps.

\vspace{3mm}

In the case of a forward contract, the actual price of the contract, including the front end protection, can be analytically computed. In particular, we have
\begin{align*}
& \E^{\Q}\left[e^{-\int_{t}^{T_\ell}r_udu}
	\indic{\tau^{i}>T_{\ell-1}}|\F_{t}\vee\mathcal{H}_t
	\right]
 = \indic{\tau^i>t}\E^{\Q}\left[
 	e^{-\int_{t}^{T_{\ell-1}}(r_u+\lambda_u)du}
	P(T_{\ell-1},T_{\ell})|\F_{t}\right]\\
& \ \ \
 = \indic{\tau^i>t}\E^{\Q}\left[
 	e^{-(Y^r_{T_{\ell -1}}-Y^r_{t})-(Y^{\lambda}_{T_{\ell -1}}-Y^{\lambda}_{t})}
	e^{r_{T_{\ell-1}}\zeta}|\F_{t}\right]
	e^{
	\int_{T_{\ell-1}}^{T_{\ell}}\int_0^{\infty}
	\left(e^{-\frac{1-e^{-\theta_r(T_{\ell}-u)}}
	{\theta_r}y}-1\right)\varphi_r(y)dydu}\\
& \ \ \
 =  \indic{\tau^i>t}e^{-\xi_r(T_{\ell-1}-t,r_t,1,\zeta)
	+\xi_{\lambda}(T_{\ell-1}-t,\lambda_t,1,0)}\\
	& \ \ \ \ \
	\times e^{
	\int_{t}^{T_{\ell-1}}
	\int_0^{\infty}\left(
	e^{-\psi^r_u(T_{\ell-1},1,1,\zeta,0)y}-1\right)
	\varphi_r(y)dydu
	+\int_{t}^{T_{\ell-1}}
	\int_0^{\infty}\left(
	e^{-\psi^{\lambda}_u(T_{\ell-1},1,0)y}-1\right)
	\varphi_{\lambda}(y)dydu}\\
	& \ \ \ \ \
	\times e^{
	\int_{T_{\ell-1}}^{T_{\ell}}\int_0^{\infty}
	\left(e^{-\frac{1-e^{-\theta_r(T_{\ell}-u)}}
	{\theta_r}y}-1\right)\varphi_r(y)dydu},
\end{align*}
and
\begin{align*}
& \E^{\Q}\left[e^{-\int_{t}^{T_\ell}r_udu}
	\indic{\tau^{i}>T_{\ell}}|\F_{t}\vee\mathcal{H}_t\right]
 = \indic{\tau^i>t}\E^{\Q}\left[
 	e^{-(Y^r_{T_{\ell -1}}-Y^r_{t})-(Y^{\lambda}_{T_{\ell -1}}-Y^{\lambda}_{t})}
	|\F_t\right]\\
& \ \ \
 = 	\indic{\tau^i>t}e^{-\xi_r(T_{\ell}-t,r_t,1,0)
	-\xi_{\lambda}(T_{\ell}-t,\lambda_t,1,0)}\\
	& \ \ \ \ \
	\times e^{\int_{t}^{T_{\ell}}
	\int_0^{\infty}\left(
	e^{-\psi^r_u(T_{\ell},1,1,0,0)y}-1\right)
	\varphi_r(y)dydu
	+\int_{t}^{T_{\ell}}
	\int_0^{\infty}\left(e^{-\psi^{\lambda}_u	
	(T_{\ell},1,0)y}-1\right)
	\varphi_{\lambda}(y)dydu},
\end{align*}
where $\zeta$ is defined as above.

Figures \ref{CDX} shows that the value of the front end protection is relatively small for this set of parameters, although, as noted for instance in \cite{BrigoMorini}, its value can be substantial for higher values of $\lambda_0$.

\vspace{3mm}

\begin{figure*}
        \centering
        \begin{subfigure}[b]{0.40\textwidth}
            \centering
            \includegraphics[width=\textwidth]{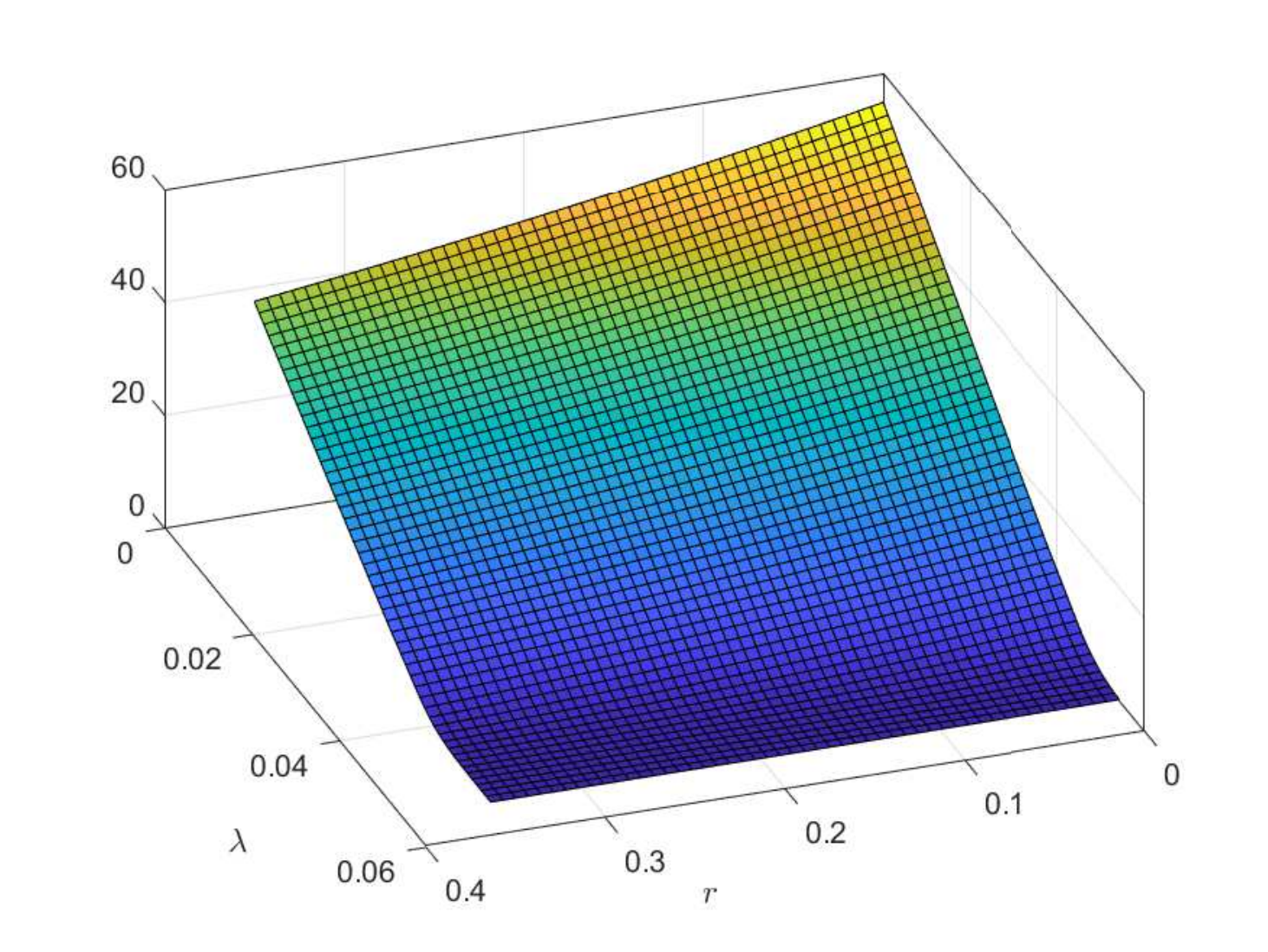}
            \caption[]%
            {{}}    
        \end{subfigure}
        \hfill
        \begin{subfigure}[b]{0.35\textwidth}  
            \centering 
            \includegraphics[width=\textwidth]{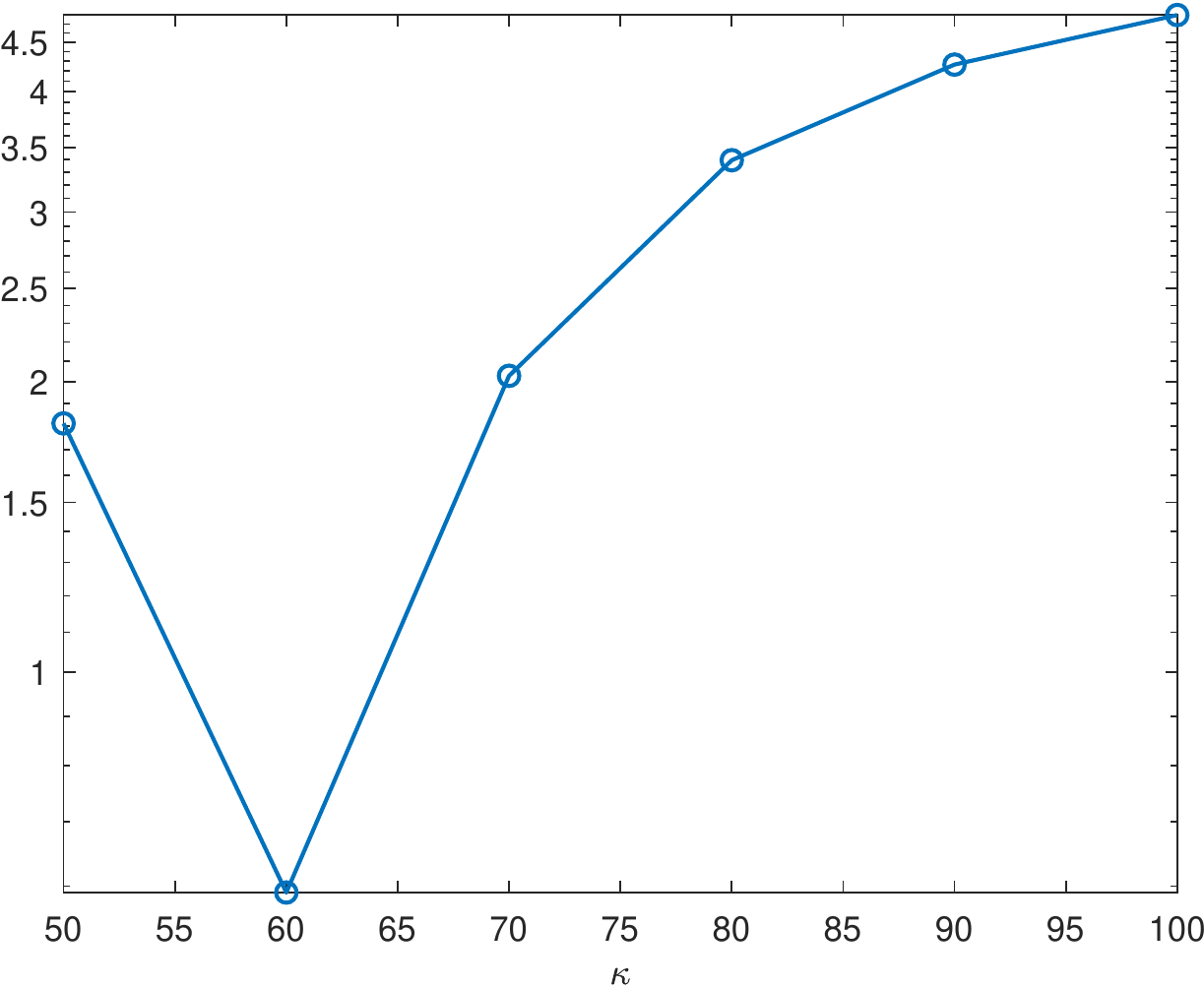}
            \caption[]%
            {{}}  
        \end{subfigure} 
        \caption{Receiver CDX option price surface for $\kappa=60$ bps and $N=50$ (left) and $\ell^{\infty}$ absolute difference with the Montecarlo generated price surface for $\kappa=50:10:100$.}
        \label{CDXOfigure}
\end{figure*}

We now turn our attention to the option contracts on a CDX index. The numerical price is shown in figure \ref{CDXOfigure} (a), while the $\ell^{\infty}$ absolute error with respect to the Montecarlo generated surface for $\kappa=50:10:100$ and for $r_0=146$ bps is shown in figure \ref{CDXOfigure} (b).

Finally, the question of convergence of the numerical method for the case of an option payoff is addressed. For $r_0=146$ bps, consider CDX spreads $\kappa$ in the range $60$ to $50$ bps. The resulting prices for various values of $N$, reported in table \ref{table:CDXO}, show that convergence up to the second decimal (in bps) is obtained already for $N=50$.

\begin{table}[H]
\begin{center}
  \begin{tabular}{ c || c | c || c }
	$N/\kappa$ 
	& \parbox{2.8cm}{\centering
		{$60$ bps}}
	& \parbox{2.8cm}{\centering
		{$50$ bps }}
	& Cpu time\\
	\hline
     50 & 53.98734 & 12.05898
        & 4.656107e+00\\
    100 & 53.98690 & 12.05998
        & 1.225736e+01\\
	150 & 53.98675 & 12.06023
		& 2.448375e+01\\
	200 & 53.98669 & 12.06029
		& 4.205160e+01\\
	250 & 53.98665 & 12.06030
		& 6.590890e+01
  \end{tabular}
\end{center}
\caption{CDXO price (in bps) and cpu time for different values of $N$ and strike price $\kappa$ bps and for $M=N_{sim}=100$.}
\label{table:CDXO}
\end{table}

\section{Comparison with Market Data}\label{Experiments}
\begin{figure*}
\centering
1\includegraphics[width=0.4\textwidth]{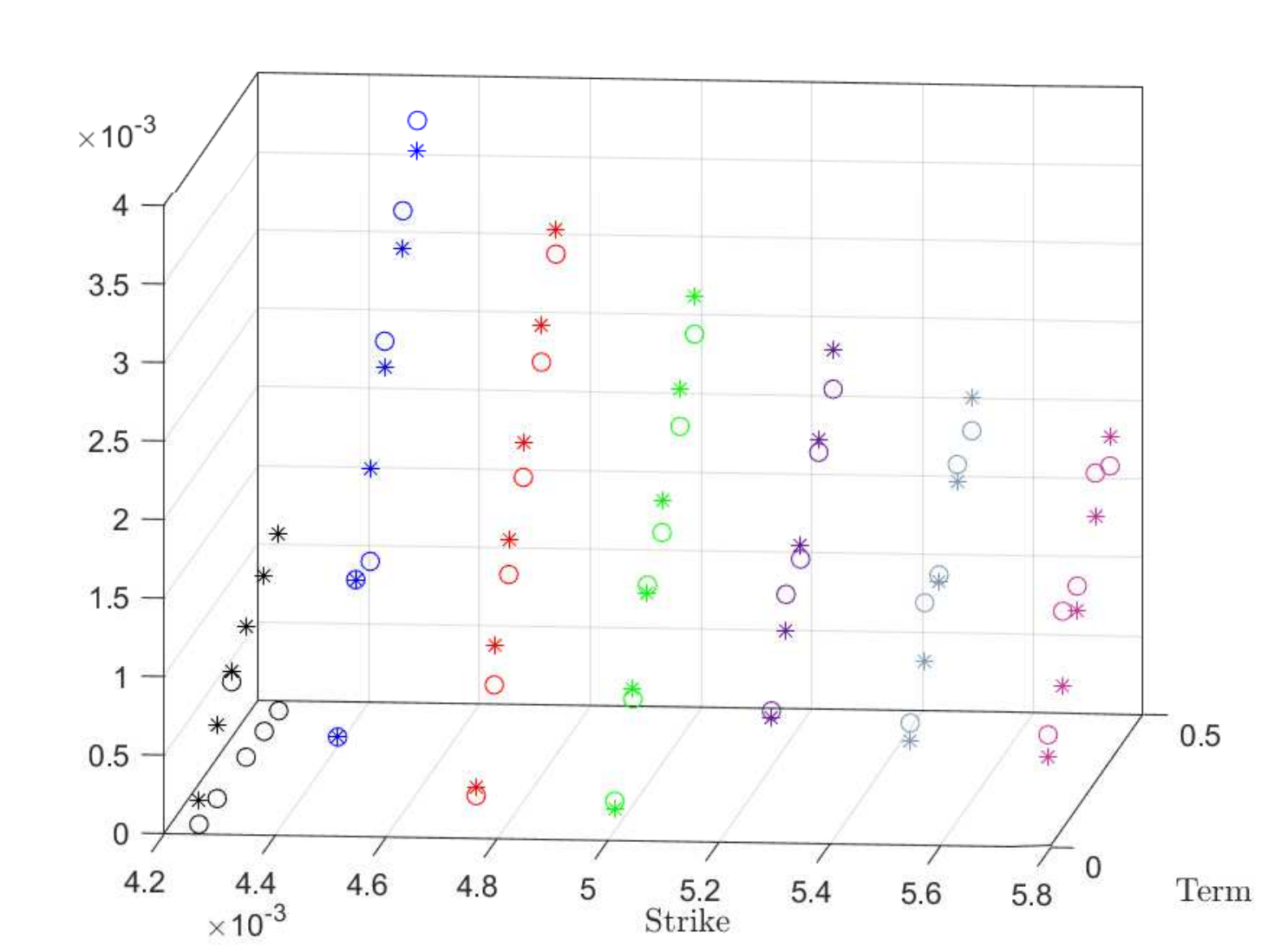}
\caption{OTM mid price (asterisk) and model price (circle) for maturities and strikes traded on 2 January 2020. Each color corresponds to one of the following strikes: $42.5$, $45$, $47.5$, $50$, $52.5$, $55$, $57.5$ (in bps). Maturities are as reported in table \ref{table:Calibration} and model prices are computed using the corresponding parameters also reported in table \ref{table:Calibration}.}
\label{CalibrationResult}
\end{figure*}

We calibrated the model to the Treasury yield curve (from \href{www.treasury.gov}{www.treasury.gov}) and to CDX option prices (provided by Morgan Stanley) as of 2 January 2020 across traded strikes and for each traded maturity. Strike prices are expressed in terms of CDX spreads, and they range from 42.5 bps up to 120 bps. Traded maturities are 13, 43, 76, 104, 139 and 167 business days. The spot CDX spread as of 2 January 2020 was 44 bps. We considered strikes that are up to 30$\%$ out of the money (OTM) for receiver and payer contracts for each available maturity.\footnote{A receiver option, i.e. an option to sell protection, is OTM if the spot spread is higher than the strike spread, while a payer option is OTM if the spot spread is lower than the strike spread. } Calibrated parameters for the short rate are the same as those considered above, while those for the default intensity are reported in table \ref{table:Calibration}. Figure \ref{CalibrationResult} compares the corresponding OTM model and market price.
\begin{table}[H]
\begin{center}
  \begin{tabular}{ c || c | c | c | c | c | c }
	\parbox{2.8cm}{\centering
		{Term (years)}}
	& $\theta_{\lambda}$
	& $\rho$
	& $c_{\lambda}$ 
	& $\gamma_{\lambda}$
	& $c_{\tau}$
	& $\gamma_{\tau}$ \\
	\hline
	0.04 & 0.1562 & 0.7869 & 20.3292 & 4.1223 
		 & 604.0000 & 3.3192 \\
	0.13 & 3.3533 & 0.1548 & 4.3178 & 6.0617 
		 & 190.0001 & 3.5298 \\
	0.21 & 2.6789 & 0.1115 & 6.1313 & 2.6983
		 & 101.2590 & 3.6123 \\
	0.29 & 0.0026 & 0.1280 & 18.7756 & 5.1836 
		 & 312.5091 & 2.5903 \\
	0.39 & 0.0010 & 0.1000 & 10.0981 & 4.4205 
		 & 818.1465 & 4.9855 \\
	0.46 & 0.0010 & 0.1000 & 82.2892 & 1.0241 
		 & 45.8397 & 8.4584 \\
  \end{tabular}
\end{center}
\caption{}
\label{table:Calibration}
\end{table}

It is also possible to compare market and model implied summary statistics (variance, skewness, kurtosis) of credit spreads for a specific maturity. In particular, as shown above, the payoff of a receiver CDX option maturing at time $T_0$ with strike spread $\overline{c}$ is given by
\begin{align*}
&\pi_{T_0}
 =\left[
 	\overline{c}\E^{\Q}[A_{T_0}|\F_{T_0}]-\delta\E^{\Q}[\Phi_{T_0}|\F_{T_0}]\right]^+,
\end{align*}
while the spot credit spread $c_{T_0}$ satisfies
\begin{align*}
c_{T_0}\E^{\Q}\left[A_{T_0}|\F_{T_0}\right]
&=\delta\E^{\Q}\left[\Phi_{T_0}|\F_{T_0}\right].
\end{align*}
Therefore, we have
\begin{align*}
&\pi_{T_0}
 =	\E^{\Q}[A_{T_0}|\F_{T_0}]\left(\overline{c}-c_{T_0}\right)^+.
\end{align*}
Taking the index annuity as numeraire,\footnote{Technically, the index annuity may be null on a set of positive measure. This happens in the case of an armageddon event, i.e. all the entities in the index default prior to the option expiration. Our assumption that such event has approximately zero probability (which is more likely the shorter the maturity of the option) ensures that the approximation error in taking the index annuity as numeraire is small enough.} and letting $\Q^A$ denote the associated probability measure, the price $u_r(t,c)$ and $u_p(t,c)$ of a receiver/payer option at time $t$ is given by
\begin{align*}
u_r(t,\overline{c})=\E^{\Q}[A_t|\F_t]\E^{\Q^A}
	[\left(\overline{c}-c_{T_0}\right)^+|\F_t], \
 u_p(t,\overline{c})=\E^{\Q}[A_t|\F_t]\E^{\Q^A}
	[\left(c_{T_0}-\overline{c}\right)^+|\F_t].	
\end{align*}
We can then use a result due to Madan and Carr (see \cite{CarrMadan2001}), according to which a twice continuously differentiable payoff function $H(c)\in \mathcal{C}^2(\R)$ can be written as
\begin{align}\label{CallPutDecomp}
H(c)
 = H(\hat{c})+(c-\hat{c})H'(\hat{c})
  + \int_{\hat{c}}^{\infty}H''(\overline{c})
  	(c-\overline{c})^+d\overline{c}
  +\int_{0}^{\hat{c}}H''(\overline{c})
  	(\overline{c}-c)^+d\overline{c},
\end{align}
where $\hat{c}\geq 0$ is arbitrary. Applying (\ref{CallPutDecomp}) to the volatility, cubic and quartic contracts defined by $(c-c_f)^2, \ (c-c_f)^3, \ (c-c_f)^4$, where $c_f=\E^{Q^A}[c_{T_0}]$ is the forward credit spread, gives the following model-free summary statistics of the spot credit spread $c_{T_0}$ under $\Q^A$:
\begin{align*}
& \E^{\Q^A}_0[(c_{T_0}-c_f)^2] 
  = 2\E^{\Q^A}_0\left[
	\int_{c_f}^{\infty}
	(c_{T_0}-\overline{c})^+d\overline{c}
	+\int_0^{c_f}
	(\overline{c}-c_{T_0})^+d\overline{c}\right], \\
& \E^{\Q^A}_0[(c_{T_0}-c_f)^3] 
 = 6\E^{\Q^A}_0\left[
	\int_{c_f}^{\infty}
	(\overline{c}-c_f)(c_{T_0}-\overline{c})^+d\overline{c}
	+\int_0^{c_f}
	(\overline{c}-c_f)(\overline{c}-c_{T_0})^+d\overline{c}\right], \\
& \E^{\Q^A}_0[(c_{T_0}-c_f)^4] 
 = 12\E^{\Q^A}_0\left[
	\int_{c_f}^{\infty}
	(\overline{c}-c_f)^2(c_{T_0}-\overline{c})^+d\overline{c}
	+\int_0^{c_f}
	(\overline{c}-c_f)^2(\overline{c}-c_{T_0})^+d\overline{c}\right].
\end{align*}
This implies, under reasonable assumptions on $c_{T_0}$,
\begin{align}
 \E^{\Q}_0[A_0]\E^{\Q^A}_0[(c_{T_0}-c_f)^2] 
 & =
    2\int_{c_f}^{\infty}u_p(0,\overline{c})d\overline{c}
    +2\int_0^{c_f}u_r(0,\overline{c})d\overline{c},
    \label{vol} \\
\E^{\Q}_0[A_0]\E^{\Q^A}_0[(c_{T_0}-c_f)^3]
 & =
    6\int_{c_f}^{\infty}(\overline{c}-c_f)u_p(0,\overline{c})d\overline{c}
    +6\int_0^{c_f}(\overline{c}-c_f)u_r(0,\overline{c})d\overline{c},
	\label{cub} \\
 \E^{\Q}_0[A_0]\E^{\Q^A}_0[(c_{T_0}-c_f)^4]
 & =
    12\int_{c_f}^{\infty}(\overline{c}-c_f)^2u_p(0,\overline{c})d\overline{c}
    +12\int_0^{c_f}(\overline{c}-c_f)^2u_r(0,\overline{c})d\overline{c},
	\label{qua}
\end{align}
Assuming that non traded deep OTM otions have zero price, one can think of (\ref{vol}), (\ref{cub}) and (\ref{qua}) as reasonable approximations of the first three moments of $c_{T_0}$, multiplied by the current value of the index annuity. Note also that calculation of $c_f$ is straightforward since, following the standard put-call parity argument, the price $f_p(t,c)$ at time $t$ of a payer forward with spread $c$ is
\begin{align*}
f_p(T_0,c_f)
& = \left[\delta\E^{\Q}[\Phi_{T_0}|\F_{T_0}]
	-c_f\E^{\Q}[A_{T_0}|\F_{T_0}]\right]^+
	-\left[c_f\E^{\Q}[A_{T_0}|\F_{T_0}]
	-\delta\E^{\Q}[\Phi_{T_0}|\F_{T_0}]
	\right]^+\\
& = u_p(T_0,c_f)-u_r(T_0,c_f)
\end{align*}
and so $f_p(0,c_f)= u_p(0,c_f)-u_r(0,c_f)$. Since 
\begin{align*}
f_p(0,0)=\E^{\Q}_0[A_0]\E^{\Q^A}_0[c_{T_0}]=\E^{\Q}_0[A_0]c_f,
\end{align*}
the no arbitrage model-free value of the annuity is
\begin{align}\label{MarketAnnuity}
\E^{\Q}_0[A_0]
=\frac{f_p(0,0)}{c_f}
\approx\frac{u_p(0,0)}{c_f}.
\end{align}
Then, the market implied spread's variance, $\mu_2$, skewness $\mu_3$ and kurtosis $\mu_4$ under $\Q^A$ are:
\begin{align*}
\mu_2
 & := \E^{\Q^A}\left[(c_{T_0}-c_f)^2\right] 
   \approx \frac{2c_f}{f_p(0,0)}\left(
   	\int_{c_f}^{\infty}u_p(0,\overline{c})d\overline{c}
    +\int_0^{c_f}u_r(0,\overline{c})d\overline{c}
    \right) \\
\mu_3
 & := \frac{\E^{\Q^A}[(c_{T_0}-c_f)^3]}
 	 {\E^{\Q^A}\left[(c_{T_0}-c_f)^2
 	 \right]^{3/2}}
   \approx \frac{6c_f}{\mu_2^{3/2}f_p(0,0)}
	\left(\int_{c_f}^{\infty}(\overline{c}-c_f)u_p(0,\overline{c})d\overline{c}
    +\int_0^{c_f}(\overline{c}-c_f)u_r(0,\overline{c})d\overline{c}\right) \\
\mu_4
 & := \frac{\E^{\Q^A}[(c_{T_0}-c_f)^4]}
 	 {\E^{\Q^A}\left[(c_{T_0}-c_f)^2
 	 \right]^{4}}
    \approx \frac{12c_f}{\mu_2^{4}f_p(0,0)}
	\left(\int_{c_f}^{\infty}(\overline{c}-c_f)^2u_p(0,\overline{c})d\overline{c}
    +12\int_0^{c_f}(\overline{c}-c_f)^2u_r(0,\overline{c})d\overline{c}\right)
\end{align*}
As shown in table \ref{table:Moments2Jan2020}, model and market implied spread statistics under $\Q^A$ are relatively close, evidencing accuracy of model (\ref{GDSRDI}) in explaining short rate and default intensity dynamics. Note in particular that model (\ref{GDSRDI}) is able to capture the positive skewness and leptokurtic feature of CDX spreads under the measure $\Q_A$.

\begin{table}[H]
\begin{center}
  \begin{tabular}{ c || c | c }
  	\multicolumn{3}{c}{\textbf{Variance}}\\
    \hline
	\parbox{1cm}{\centering
		{Term}}
	& \parbox{2.8cm}{\centering
		{Market Implied}}
	& \parbox{2.8cm}{\centering
		{Model Implied}}\\
	\hline
	0.04 & 1.054514e+01 & 7.459700e+00\\
	0.13 & 5.865457e+01 & 6.210797e+01\\
	0.21 & 1.100104e+02 & 2.152011e+02\\
	0.29 & 1.812481e+02 & 2.623242e+02\\
	0.39 & 2.722237e+02 & 5.018296e+02\\
	0.46 & 3.496845e+02 & 2.280879e+02\\
	\hline
  \end{tabular}
  \begin{tabular}{ c || c | c }
    \multicolumn{3}{c}{\textbf{Skewness}}\\
  	\hline
	\parbox{1cm}{\centering
		{Term}}
	& \parbox{2.8cm}{\centering
		{Market Implied}}
	& \parbox{2.8cm}{\centering
		{Model Implied}}\\
	\hline
	0.04 & 8.599558e-01 & 2.520823e-01\\
	0.13 & 2.463146e+00 & 2.974330e+00\\
	0.21 & 2.660094e+00 & 3.507166e+00\\
	0.29 & 3.221129e+00 & 3.992535e+00\\
	0.39 & 3.144701e+00 & 3.809120e+00\\
	0.46 & 2.896963e+00 & 2.457448e+00\\
	\hline
  \end{tabular}
  \begin{tabular}{ c || c | c }
  \multicolumn{3}{c}{\textbf{Kurtosis}}\\
  \hline
	\parbox{1cm}{\centering
		{Term}}
	& \parbox{2.8cm}{\centering
		{Market Implied}}
	& \parbox{2.8cm}{\centering
		{Model Implied}}\\
	\hline
	0.04 & 2.308373e+00 & 2.732897e+00\\
	0.13 & 9.452965e+00 & 1.140996e+01\\
	0.21 & 1.066158e+01 & 1.230041e+01\\
	0.29 & 1.566200e+01 & 1.788540e+01\\
	0.39 & 1.458586e+01 & 1.487553e+01\\
	0.46 & 1.212721e+01 & 1.028359e+01\\
  \end{tabular}
\end{center}
\caption{CDX spread statistics for maturities traded on 2 January 2020.}
\label{table:Moments2Jan2020}
\end{table}

It is also worth noting that, under (\ref{GDSRDI}),
\begin{align}\label{ModelImpliedStatistics}
\E^{\Q^A}_0[H(c_{T_0})]
& = \frac{
	\E^{\Q}_0\left[e^{-\int_0^{T_0}r_udu}
	\sum_{\ell=1}^M(T_\ell-T_{\ell-1})
	\E^{\Q}\left[e^{-\int_{T_0}^{T_{\ell}}(r_u
		+\lambda_u)du}|\F_{T_0}\right]
	H(c_{T_0})\right]}
	{\sum_{\ell=1}^M(T_\ell-T_{\ell-1})
  	\E^{\Q}_0\left[e^{-\int_{0}^{T_{\ell}}(r_u
		+\lambda_u)du}\right]},
\end{align}
which can be computed via montecarlo simulation. The resulting model implied statistics are however far from those computed above, showing the relevance of prices of deep OTM options.

\begin{figure*}
        \centering
        \begin{subfigure}[b]{0.32\textwidth}
            \centering
            \includegraphics[width=\textwidth]{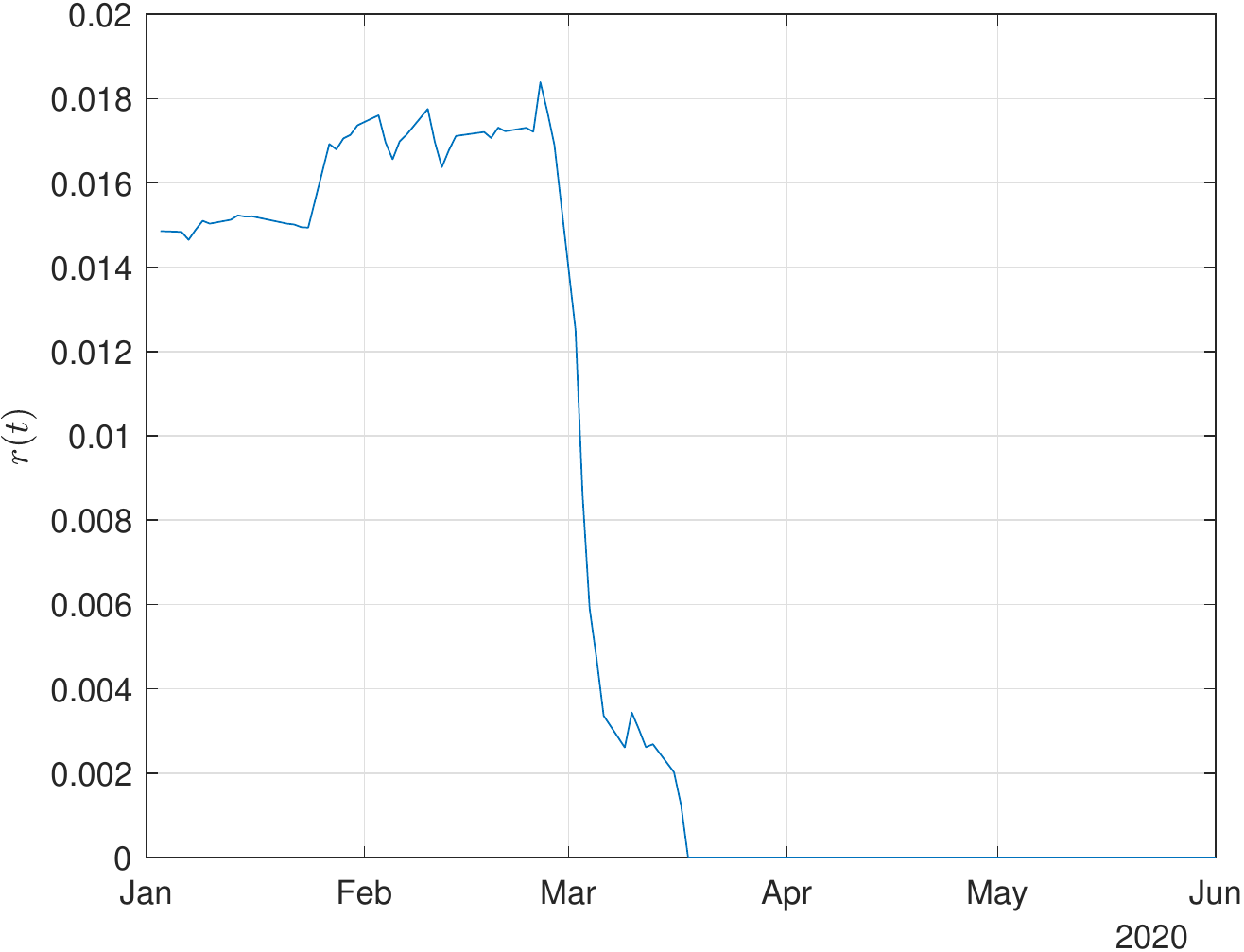}
            \caption[]%
            {{}}    
        \end{subfigure}
        \hfill
        \begin{subfigure}[b]{0.32\textwidth}  
            \centering 
            \includegraphics[width=\textwidth]{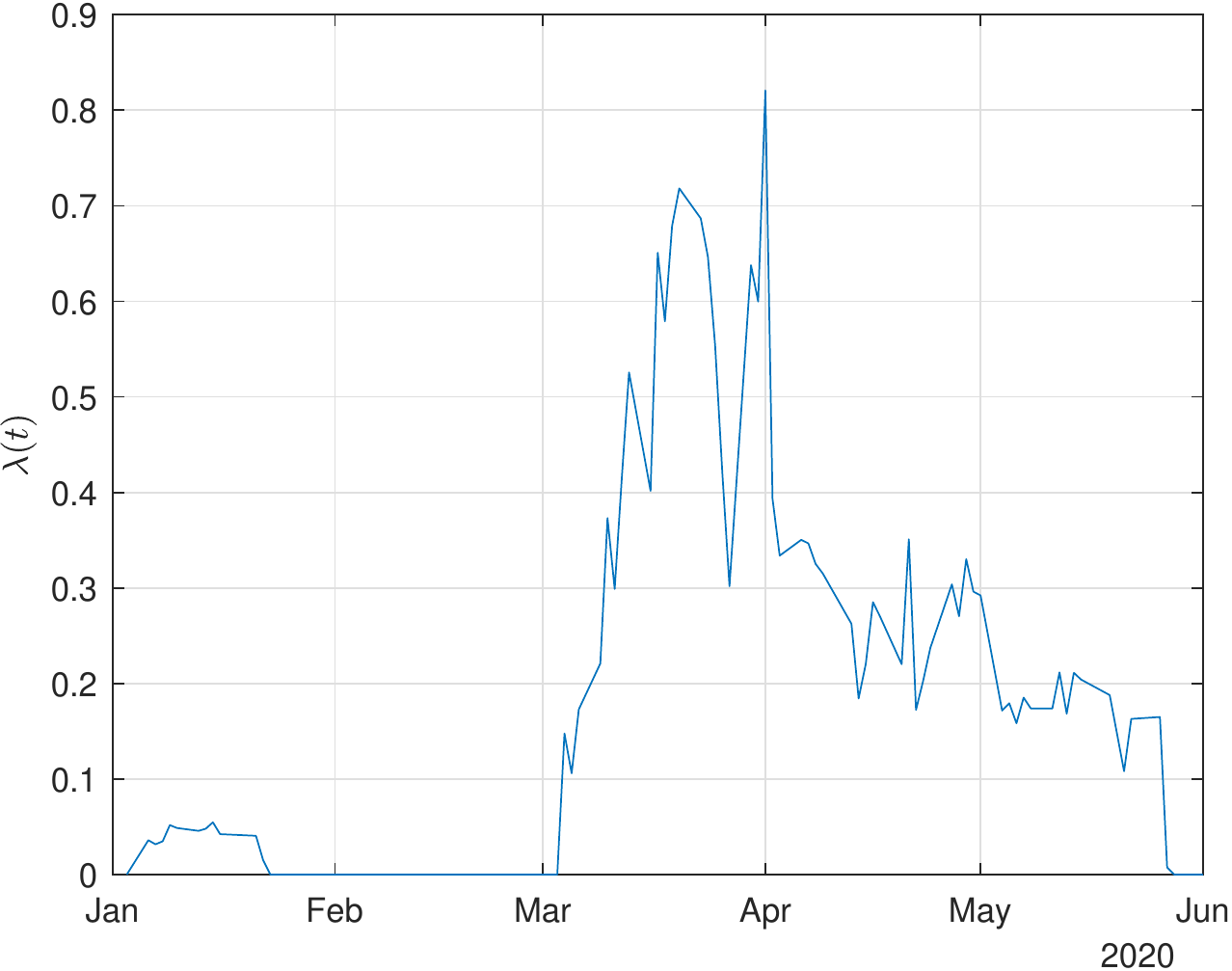}
            \caption[]%
            {{}}  
        \end{subfigure} 
        \begin{subfigure}[b]{0.32\textwidth}  
            \centering 
            \includegraphics[width=\textwidth]{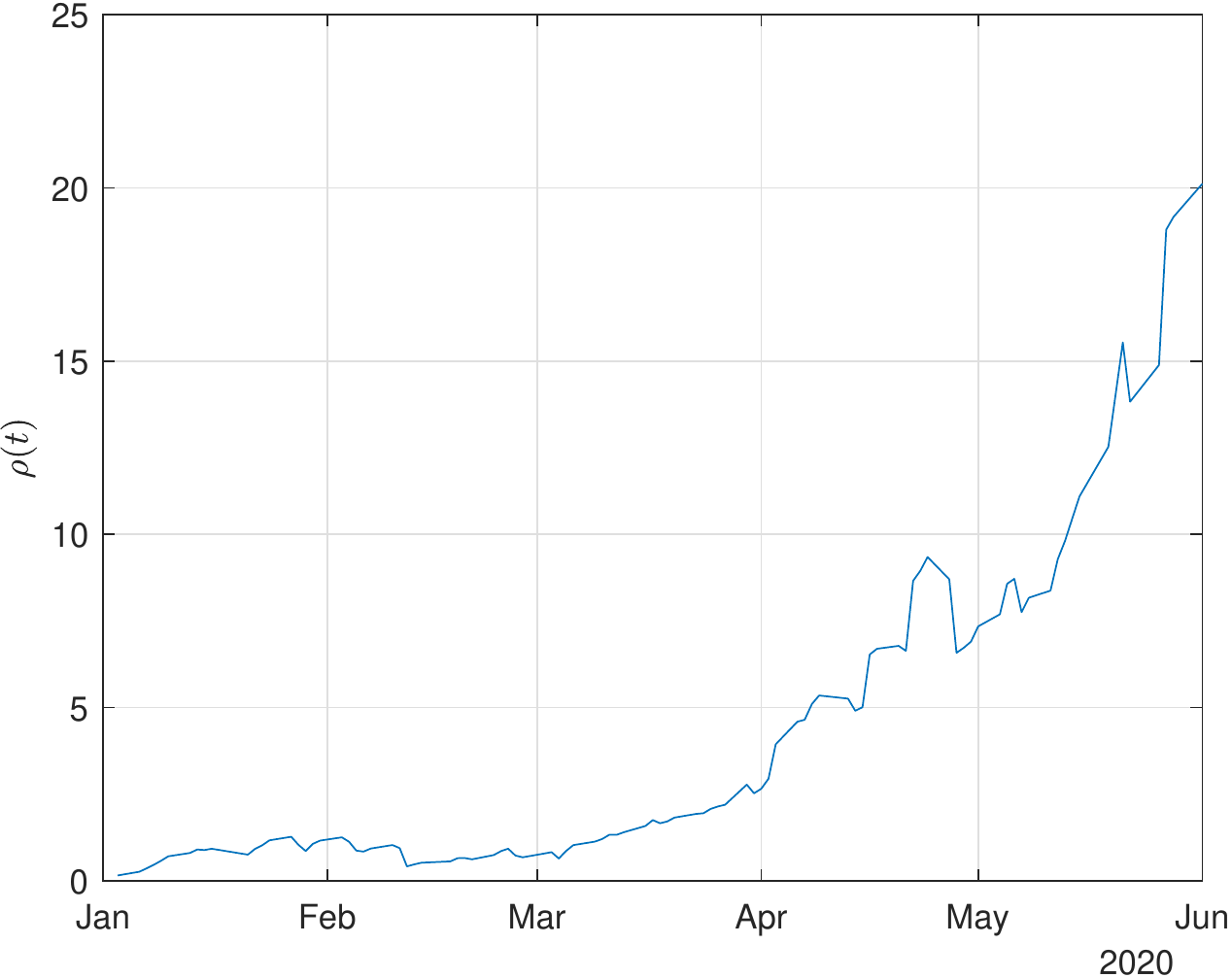}
            \caption[]%
            {{}}  
        \end{subfigure} 
        \caption{Daily realized short rate, hazard rate and parameter $\rho$.}
\label{param}
\end{figure*}

Finally, table \ref{table:Corrcoeff} shows model and market implied moments correlation between 2 January 2020 and 5 June 2020 and figure \ref{param} the daily realized short rate, default intensity and parameter $\rho$.\footnote{Calibration was performed each day via Nelder-Mead algorithm starting at previous day optimal parameters.}

\begin{table}[H]
\begin{center}
  \begin{tabular}{ c | c }
	\parbox{2.8cm}{\centering
		{Statistics}}
	& Correlation Coefficient \\
	\hline
	Variance & 0.8892 \\
	Skewness & 0.3238\\
	Kurtosis & 0.4759
  \end{tabular}
\end{center}
\caption{Correlation of time series of market and model implied spread statistics.}
\label{table:Corrcoeff}
\end{table}

\begin{figure*}
        \centering
        \begin{subfigure}[b]{0.32\textwidth}
            \centering
            \includegraphics[width=\textwidth]{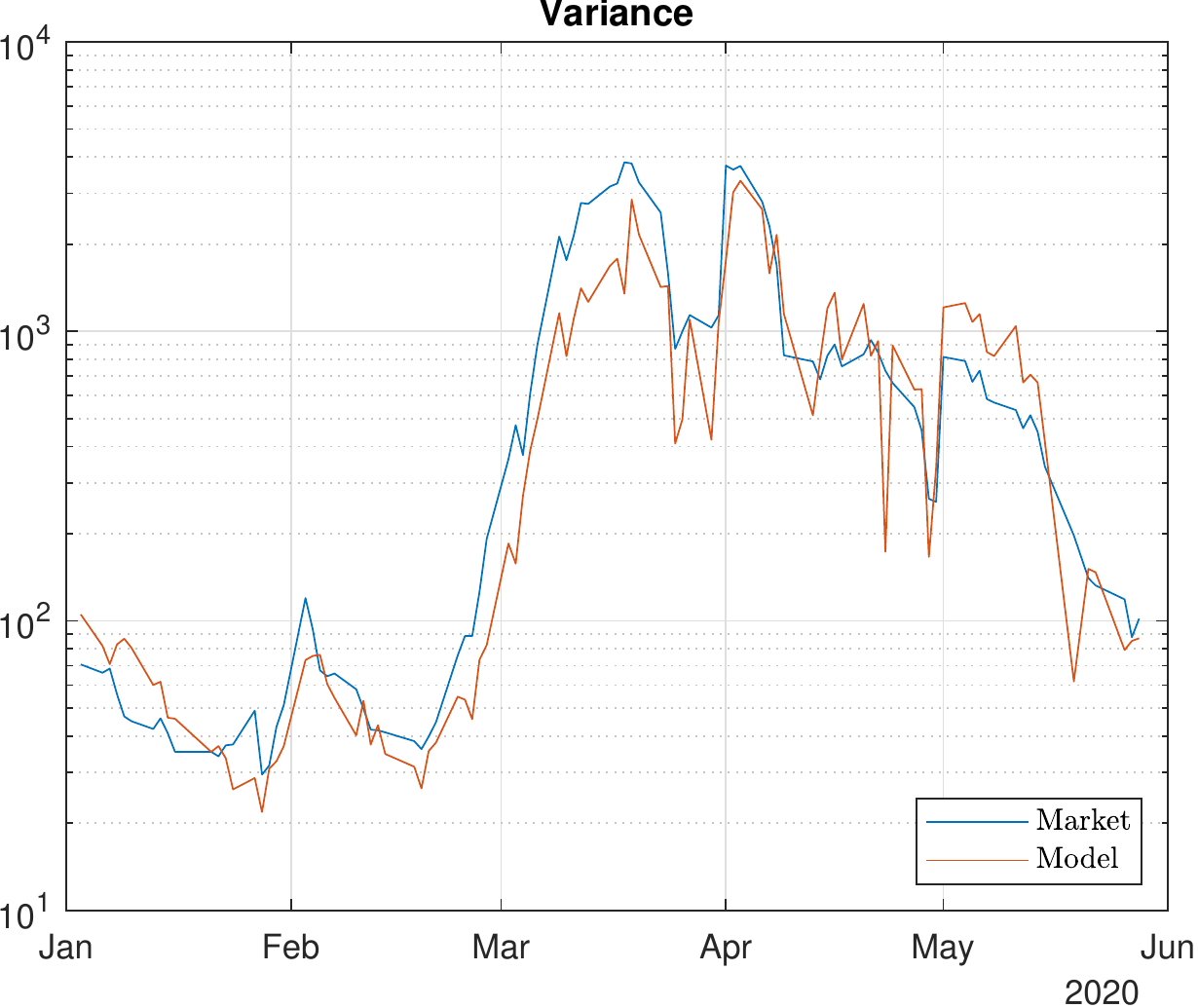}
            \caption[]%
            {{}}    
        \end{subfigure}
        \begin{subfigure}[b]{0.32\textwidth}
            \centering
            \includegraphics[width=\textwidth]{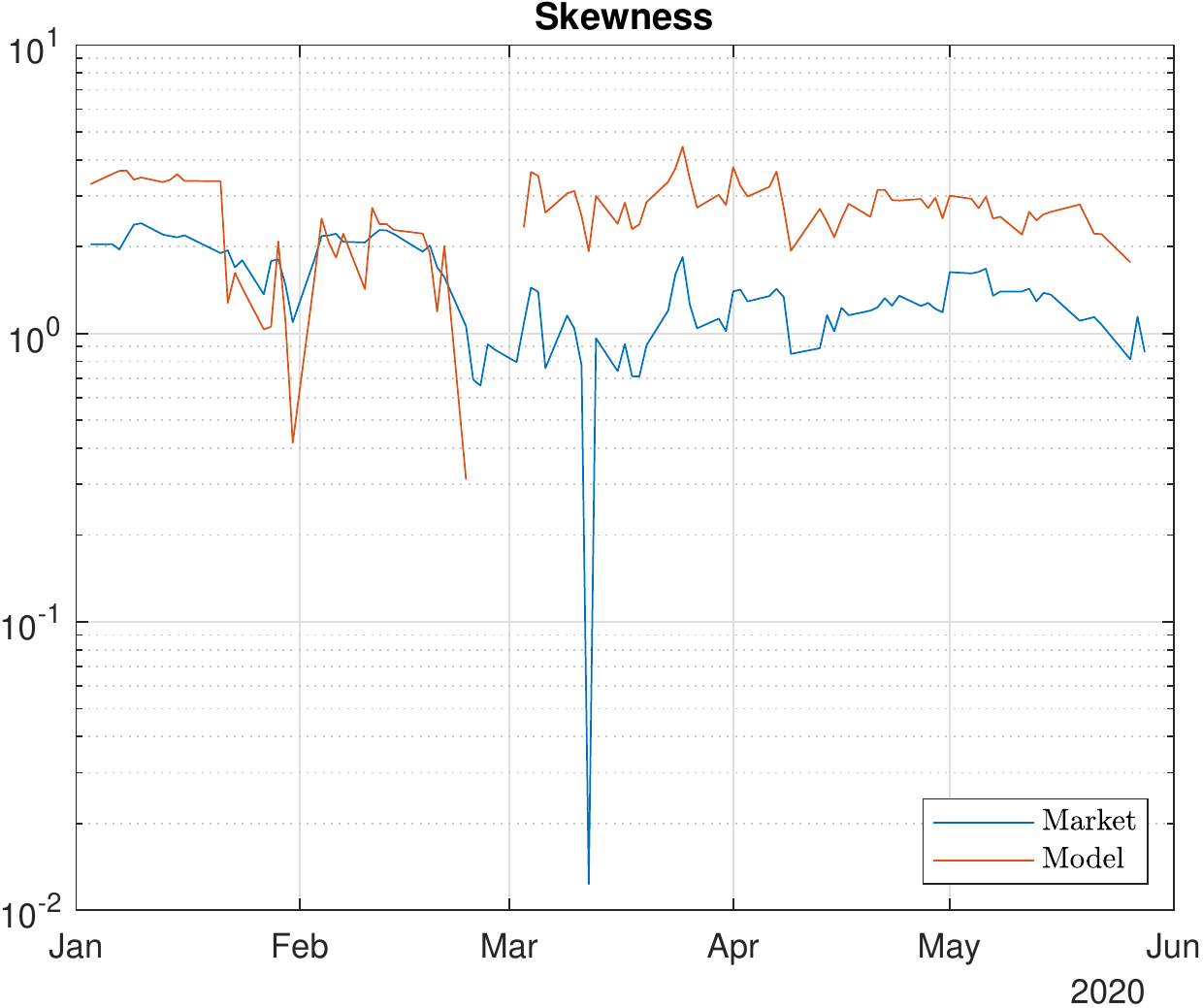}
            \caption[]%
            {{}}    
        \end{subfigure}
        \begin{subfigure}[b]{0.32\textwidth}
            \centering
            \includegraphics[width=\textwidth]{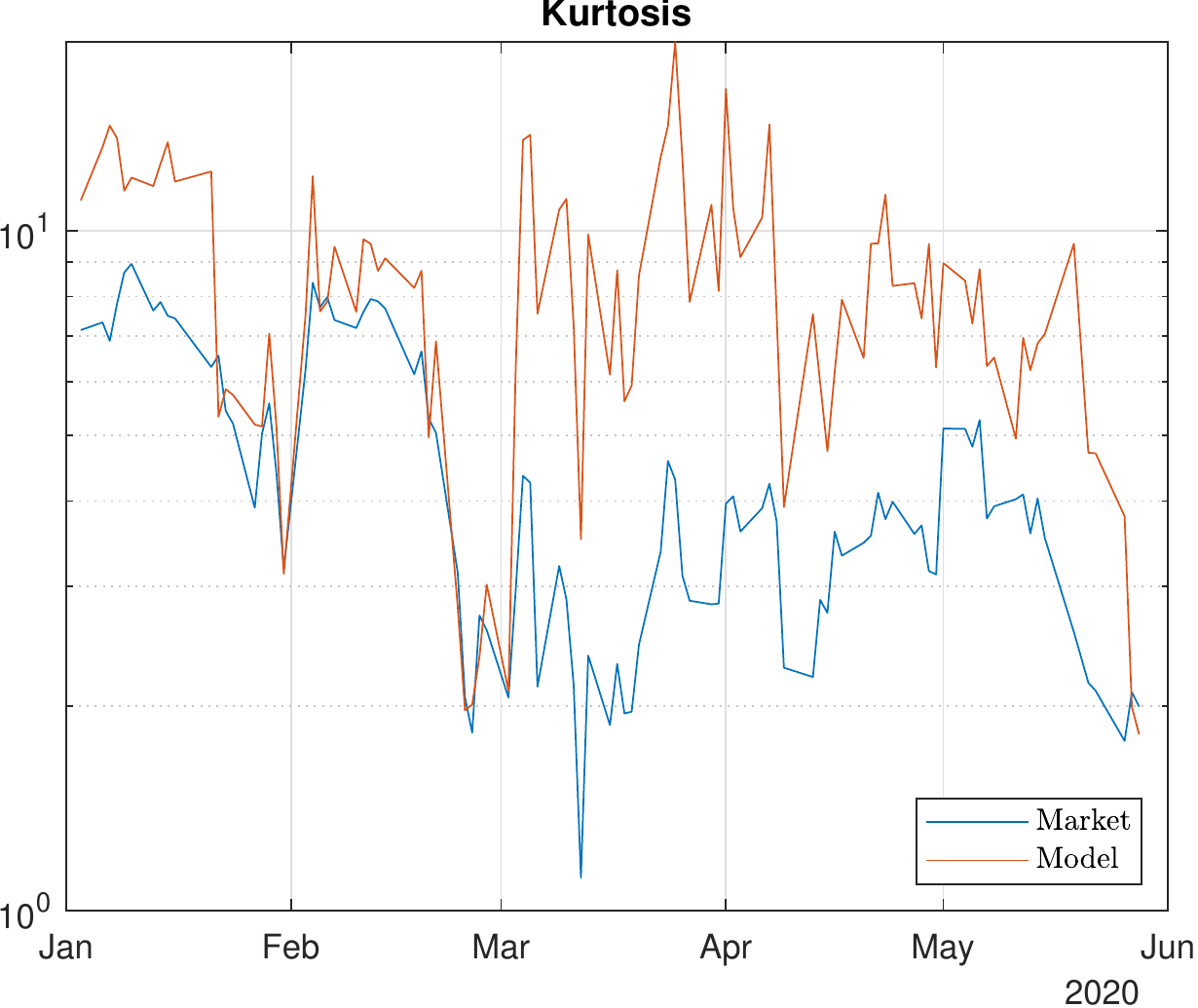}
            \caption[]%
            {{}}    
        \end{subfigure}
        \caption{Daily market implied and model statistics under $\Q^A$.}
\label{StatsFig}
\end{figure*}

\section{Conclusions}

We introduced a pure-jump dynamics for the simultaneous modelling of the short rate and default intensity of a pool of entities with similar credit qualities, with the former being a gamma process and the latter also a gamma process but subordinated to another (independent) gamma process. We tested a simple finite difference scheme for the valuation PIDE for forward and option contracts on derivatives determined by short rate and default intensity, and showed that the numerical solution approximates the exact one or a simulated one with reasonable margin of errors. We calibrated the model to the CDX option price surface. For January 2 2020, the calibration error is generally low, but it can be substantial and especially as maturity increases. Finally, we derived a market implied formula for variance, skewness and kurtosis of the credit spread under the Annuity measure, and reported that market and model implied statistics over the year 2020 are of similar magnitude and positively correlated.

\section{Aknowledgement}
This paper is a revised version of a chapter of the author's doctoral dissertation, conducted under the supervision of Prof. Dilip B. Madan at the University of Maryland, College Park. I am also grateful to an anonymous referee for carefully reviewing this manuscript. 

\section{Appendix: The Finite Difference Scheme for the Valuation PIDE}
In our finite difference approximation, we treat the integral term fully explicitly. Consider the following mesh on the region $[0,T]\times [0,r_{\max}]\times[0,\lambda_{\max}]$:
\begin{equation*}
D =
\begin{cases}
t_j=j\Delta t;\ \Delta t=\frac{T}{M}; \ j=0,1,...,M\\
r_i=i\Delta r;\ \Delta r=\frac{r_{\max}}{N}; \ i=0,1,...,N\\
\lambda_k=k\Delta \lambda;\ 
	\Delta \lambda=\frac{\lambda_{\max}}{L}; \ k=0,1,...,L
\end{cases}
\end{equation*}
We pick $\lambda_{\max}=\rho r_{\max}$ and $L=N$. We denote by $(t_j,r_i,\lambda_k)\in\R^3_+$ the grid points in $D$, and let $u^j_{i,k}=u(t_j,r_i,\lambda_k)$. Assuming that the $(N+1)^2$ values $u^j_{i,k}$ are known for fixed $t_j$, we need to construct the difference equation for each point $(t_{j+1},r_i,\lambda_k)$. Space and time derivatives are approximated using central and forward differences respectively, i.e.
\begin{align*}
& u_{r}(t_{j+1},r_i,\lambda_k)
  =\frac{u^{j+1}_{i+1,k}-u^{j+1}_{i-1,k}}{2\Delta r}+O(\Delta r^2), \\
  u_{\lambda}(t_{j+1},r_i,\lambda_k)
& =\frac{u^{j+1}_{i,k+1}-u^{j+1}_{i,k-1}}
 	{2\Delta \lambda}+O(\Delta \lambda^2),\\
& u_{t}(t_{j+1},r_i,\lambda_k)
  =\frac{u^{j+1}_{i,k}-u^{j}_{i,k}}
 	{\Delta t}+O(\Delta t^2).
\end{align*}
We then obtain the following equation for the point $(t_{j+1},r_i,\lambda_k)$:
\begin{align*}
\frac{u^{j+1}_{i,k}-u^{j}_{i,k}}{\Delta t}
	+r_iu_{i,k}^{j+1}-\alpha_1
	\frac{u^{j+1}_{i+1,k}-u^{j+1}_{i-1,k}}{2\Delta r}
	-\alpha_2
	\frac{u^{j+1}_{i,k+1}-u^{j+1}_{i,k-1}}{2\Delta \lambda}\\
\approx
	\int_{0}^{\infty}
	\mathcal{D}_u^{t_j,r_i,\lambda_k}(y_r,0)
	\varphi_r(y_r)dy_r
	+\int_{0}^{\infty}
	\mathcal{D}_u^{t_j,r_i,\lambda_k}(0,y_{\lambda})
	\varphi_{\lambda}(y_{\lambda})dy_{\lambda},
\end{align*}
Equivalently, we have
\begin{align}
& u^{j+1}_{i,k}(1+\Delta tr_i)
-\frac{\Delta t\alpha_1}{2\Delta r}
	\left(u^{j+1}_{i+1,k}-u^{j+1}_{i-1,k}\right)
	-\frac{\Delta t\alpha_2}{2\Delta \lambda}
	\left(u^{j+1}_{i,k+1}-u^{j+1}_{i,k-1}\right)\notag\\
& \ \ \
\approx
	u^{j}_{i,k}+\int_{0}^{\infty}
	\mathcal{D}_u^{t_j,r_i,\lambda_k}(y_r,0)
	\varphi_r(y_r)dy_r
	+\int_{0}^{\infty}
	\mathcal{D}_u^{t_j,r_i,\lambda_k}(0,y_{\lambda})
	\varphi_{\lambda}(y_{\lambda})dy_{\lambda}. \label{GDSRDIEqn}
\end{align}

The integral terms in (\ref{GDSRDIEqn}) can be treated easily via Montecarlo simulation. For instance, at the grid point $(t_j,r_i,\lambda_k)$, and given $N_{sim}$ iid exponentials $\{Y_s^r\}_{s=1,...,N_{sim}}$ with parameter $c_r$, we have
\begin{align*}
\int_{0}^{\infty}
	\mathcal{D}_u^{t_j,r_i,\lambda_k}(y_r,\rho y_r)
	\varphi_r(y_r)dy_r
\approx \frac{1}{N_{sim}}\sum_{s=1}^{N_{sim}}
	\mathcal{D}_u^{t_j,r_i,\lambda_k}(Y_s^r,\rho Y_s^r)
	\frac{\gamma_r}{c_rY_s^r}.
\end{align*}
For $s=1,...,N_{sim}$, we compute $\mathcal{D}_u^{t_j,r_i,\lambda_k}(Y_s^r,\rho Y_s^{r})$ and $\mathcal{D}_u^{t_j,r_i,\lambda_k}(0,Y_s^{\lambda})$ by linearly interpolating $u^j$, and obtain the following difference equation
\begin{align}
-S_{i,k}u_{i,k-1}^{j+1}
-W_{i,k}u_{i-1,k}^{j+1}+C_{i,k}u_{i,k}^{j+1}
	-E_{i,k}u_{i+1,k}^{j+1}
-N_{i,k}u_{i,k+1}^{j+1}
= u^j_{i,k}+\Delta tR^j_{i,k},
\label{GDSRDIDiffEqn}
\end{align}
where
\begin{align*}
S_{i,k}
& = -\frac{\Delta t\alpha_2}{2\Delta\lambda},
  \ W_{i,k}
  = -\frac{\Delta t\alpha_1}{2\Delta r},
  \ E_{i,k}
  = \frac{\Delta t\alpha_1}{2\Delta r},
  \ C_{i,k} = 1+\Delta tr_i,
  \ N_{i,k}
  = \frac{\Delta t\alpha_2}{2\Delta\lambda}, \\
R^j_{i,k}
& = \frac{1}{N_{sim}}\sum_{s=1}^{N_{sim}}
	\mathcal{D}_u^{t_j,r_i,\lambda_k}(Y_s^r,\rho Y_s^r)
	\frac{\gamma_r}{c_rY_s^r}
 + \frac{1}{N_{sim}}\sum_{s=1}^{N_{sim}}
	\mathcal{D}_u^{t_j,r_i,\lambda_k}(Y_s^{\lambda},0)
	\frac{\gamma_{\lambda}}{c_{\lambda}Y_s^{\lambda}}.
\end{align*}
\subsubsection*{Implementation of Boundary Conditions}
We impose homogeneous Neumann boundary conditions:
\begin{align}\label{NeumannGDSDRI}
u_{\lambda\lambda}(t,r,\lambda_L)=0
-u_{\lambda\lambda}(t,r,0)=0, \
-u_{rr}(t,0,\lambda)=0, \
u_{rr}(t,r_N,\lambda)=0
\end{align}
for each time $t\in[0,T]$. We thus solve
\begin{equation}
\begin{cases}
u_t+ru
	-\nabla u \cdot \alpha
	=\int_{0}^{\infty}
	\mathcal{D}_u^{t,r,\lambda}(y_r,\rho y_r)
	\varphi_r(y_r)dy_r
	+\int_{0}^{\infty}
	\mathcal{D}_u^{t,r,\lambda}(0,y_{\lambda})
	\varphi_{\lambda}(y_{\lambda})dy_{\lambda}\\
u_{\lambda\lambda}(t,r,\lambda_L)=0, \
u_{\lambda\lambda}(t,r,0)=0, \
u_{rr}(t,0,\lambda)=0, \
u_{rr}(t,r_N,\lambda)=0
\end{cases}\label{BC_S}
\end{equation}
We implement (\ref{NeumannGDSDRI}) at the points $(t_j,r_1,\lambda_k)$, $(t_j,r_{N-1},\lambda_k)$, $(t_j,r_i,\lambda_1)$, i.e. we set
\begin{align*}
u^{j+1}_{0,k}
 = 2u^{j+1}_{1,k}-u^{j+1}_{2,k}, \
u^{j+1}_{N,k}
 = 2u^{j+1}_{N-1,k}-u^{j+1}_{N-2,k}, \
u^{j+1}_{i,L+1}
 = 2u^{j+1}_{i,L}-u^{j+1}_{i,L-1}.
u^{j+1}_{i,0}
 = 2u^{j+1}_{i,1}-u^{j+1}_{i,2}.
\end{align*}

\bibliographystyle{authordate1}
\bibliography{mybib_2011_pl,mybib_2011_pl_1,mybib2,mybib2_1}

\end{document}